\documentclass[a4paper,12pt,final,fleqn,epsfig]{article}

\usepackage{natbib}
\usepackage{color, colortbl}
\usepackage{setspace}
\usepackage{amssymb}
\usepackage{graphicx}
\usepackage{pdflscape}
\usepackage{lscape}
\usepackage{xcolor}
\usepackage{todonotes}
\definecolor{darkblue}{rgb}{0,0,0.5}
\definecolor{Gray}{gray}{0.9}
\usepackage{pstricks,pst-plot}
\usepackage{booktabs}
\usepackage{threeparttable}
\usepackage{todonotes}
\usepackage{subfloat}
\usepackage{subfig}
\usepackage{amsmath,amssymb,latexsym,tabularx}
\usepackage[pdfborder=0, colorlinks=true, linkcolor=darkblue, citecolor=darkblue, urlcolor=darkblue]{hyperref}

\usepackage[ansinew]{inputenc}

\usepackage{courier}

\usepackage{subfig}

\usepackage[format=hang]{caption} 

\PassOptionsToPackage{gray}{xcolor} 
\usepackage{etex} 
\usepackage{tikz, pgf}

\usepackage[english]{babel}
\usepackage{blindtext}

\begin{document}
\begin{titlepage}
\title{\scshape{\Large Exposure to World War II and Its Labor Market Consequences over the Life Cycle\thanks{\small We would like to thank the editor of the Journal of Economic History, four anonymous referees as well as seminar and conference participants at Bielefeld University, the TU Dortmund University, the University of Bayreuth, the University of Bonn, the University of Regensburg, the 2022 Workshop on the Future of Labor in Hamburg, the 2023 Workshop of the Ausschuss f\"{u}r Bev\"{o}lkerungs\"{o}konomik, the Simposio de la Asociaci\'on Espa\~nola de Econom\'ia 2023, the 2024 Workshop of the Ausschuss f\"{u}r Wirtschaftsgeschichte, and the 2025 CEPR Applied Micro-Economic History Workshop for their helpful comments and suggestions. We are also grateful to Herbert Obinger for providing us with cross-country data on social expenditure for war victims. Jan Stuhler gratefully acknowledges financial support from MICIU/AEI (RYC2019027614-I, CEX2021-001181-M). This manuscript has been accepted at the Journal of Economic History and is made available under the
CC BY-NC-ND license.}}}

\author{Sebastian T. Braun\footnote{\small University of Bayreuth, CReAM, and IZA (\mbox{sebastian.braun@uni-bayreuth.de})}
\hspace*{1.0cm} Jan Stuhler\footnote{\small Universidad Carlos III de Madrid, CReAM, and IZA (\mbox{jan.stuhler@uc3m.es})}} 

\date{June 27, 2025}
\maketitle
\centering

\vspace{-1cm}
\thispagestyle{empty}
\begin{abstract}
\begin{singlespace}
With 70 million dead, World War II remains the most devastating conflict in history. Among the survivors, millions were displaced, returned maimed from the battlefield, or endured years of captivity. We examine the effects of such war exposures on labor market careers, showing that they often become apparent only at certain life stages. While war injuries reduced employment in old age, former prisoners of war prolonged their time in the workforce before retiring. Many displaced workers, especially women, never returned to employment. These responses align with standard life-cycle theory and thus likely hold relevance for other conflicts.

\end{singlespace}

\begin{itemize}
\begin{singlespace}
\item[] \textbf{JEL Code}: J24, J26, N34 
\item[] \textbf{Keywords}: World War II; labor market careers; war injuries; prisoners of war; displacement; life-cycle models
\end{singlespace}  
\end{itemize}

\end{abstract}

\thispagestyle{empty}
\end{titlepage}

\newpage
\section{Introduction}

World War II (WWII) remains the deadliest conflict in history. In Europe alone, some 39 million people died, and many more were wounded. Millions became prisoners of war (POWs). Nazi Germany captured 5.7 million Soviet soldiers, more than half of whom died in captivity. At the same time, about 11 million German soldiers, or more than one in two, became POWs \citep{Ratza1974}, and 5.2 million were injured in the war \citep{Mueller2016}. Many civilians were killed or forcibly uprooted. 
After the war, at least twelve million Germans were displaced from Eastern Europe in one of the largest population transfers in history. What are the long-term consequences of such traumatic events for the labor market careers of the survivors? 

Using individual-level data from West Germany, we study three of the most common consequences of war: battlefield injuries, imprisonment, and displacement. We show that a life-cycle perspective is crucial, as the labor market effects of individual war experiences become visible only at certain life stages and often long after the war has ended. 
Moreover, the same shock can have very different effects depending on the survivors' age-at-exposure. 
Standard life-cycle theory can rationalize these patterns, as we show in a Ben-Porath model with endogenous retirement decisions \citep{Hazan2009}.
Consequently, policies aimed at mitigating the economic effects of war shocks should be designed with a life-cycle framework in mind. Such policies were central after WWII when many belligerent countries, including Germany and the US, spent between 10 and 35\% of total social spending on war victims \citep{Obinger2018}. 

Our study relies mainly on nationally representative life course data for the birth cohort 1919-21 from the German Life History Study (GHS), which offers two major merits. 
First, the GHS captures complete educational, employment, and family histories. This feature allows us to construct life-cycle profiles of respondents. Second, the survey asked individuals directly about their wartime experiences, such as their time as a POW or their medical history during the war. Thus, unlike most other studies, we focus on \textit{individual} war experiences rather than regional exposure to combat or bombing.

We use this data to compare the life-cycle profiles of male survivors born 1919-21 who were wounded (30\% of the respondents in our sample) or imprisoned for more than six months (47\%) to other WWII veterans. We also compare the life-cycle profiles of those who were displaced (21\%) with those who were not. In support of our empirical strategy, we show that neither family nor own characteristics predict individual exposure, in line with historical studies showing that young soldiers of all backgrounds suffered similarly from battlefield injuries or captivity. 

We find that the effect of individual war experiences often becomes apparent only at certain stages of life. For example, battlefield injuries reduced the lifetime employment of veterans by about one year, but had no effect on employment in middle age; instead, the effect is entirely explained by early retirement, as war injuries reduced the employment rate at age 60 by 14.7 percentage points (pp). In contrast, former POWs postponed retirement and their employment rate at age 60 was 8 pp higher than that of non-incarcerated soldiers. The employment penalty for displaced men, on the other hand, is greatest in their late twenties, shortly after displacement, and disappears by age 35. Like former POWs, displaced men experience lower occupational success throughout their careers.

We complement the GHS with microdata from the 1970 census to show that the effect of displacement depends critically on the age of those affected. The loss of human capital was greatest for those born around 1930, who suffered from the turbulence of displacement during the transition from school to vocational training. In contrast, children displaced before entering school acquired \textit{more} schooling than those who were not displaced. Labor market effects also varied widely by age and gender. For older workers, displacement often led to an immediate exit from the labor force. The total employment loss was greatest for those displaced around age 50, who still had a longer career ahead of them. The loss was particularly severe for women: in some age groups, less than half of the women employed before the war returned to work after displacement.

Finally, we show that these findings are broadly in line with a Ben-Porath model with endogenous retirement decisions \citep{Hazan2009}. By increasing the disutility of work, war injuries accelerate entries into retirement but are less likely to affect employment during the early or mid-career stages. Imprisonment implies a reduction in an individual's productive work span, reducing the incentives to invest in education (as the benefits accrue over a shorter period) and potentially delaying retirement (as former POWs seek to make up for lost lifetime earnings). And displacement triggers both substitution and income effects, the relative magnitude of which depends on age at the time of displacement.

Given that they explain our findings well, life-cycle models might also help to understand labor market trajectories after other wars. Of course, the exact magnitudes of these effects also depend on the design of the pension system, where in Germany disability benefits were an important pathway to retirement, and on the availability of compensation for war victims, which in Germany was contingent on severe health impairments caused by the war.

\paragraph{Related Literature.} Recent studies on the long-term individual economic impact of WWII show that exposure to warfare negatively affected education, health, and labor market outcomes \citep{Ichino2004,AkbulutYuksel2014,Kesternich2014,AkbulutYuksel2022}.\footnote{A related literature examines the impact of male mobilization for WWII and military deaths on female labor force participation. \citep[e.g.,][]{Goldin1991,Acemoglu2004,Goldin2013,Jaworski2014,Rose2018}.} For identification, the extant literature often exploits between-country variation in combat 
exposure or regional within-country variation in destruction. Hunger early in life is one key channel through which WWII had adverse long-term effects on survivors' well-being \citep[e.g.][]{Juerges2013,Kesternich2015,Mink2020}. Moreover, evidence from Germany \citep{Bauer2013} and Finland \citep{Sarvimaeki2022} shows that displacement resulted in long-term income loss, except for those who originated from agricultural areas. However, the loss of land or property might encourage investment in education, and the descendants of Polish WWII migrants are better educated today \citep{Becker2020}. WWII military service increased the educational attainment of veterans \citep{Bound2002, Cousley2017} but had little effect on earnings \citep{Angrist1994} and occupational attainment later in life \citep{Maas1999}. 

The core innovation of our study is the life-cycle perspective. We show that the labor market consequences of individual war shocks often become visible only at certain stages of life, making a life-cycle perspective essential. This result is consistent with previous work on the impact of military service, which emphasizes the importance of a long-term perspective \citep{Bedard2006,Angrist2011}.\footnote{Similarly, \cite{Kesternich2020} show that the impact of war-related sex ratio imbalances on fertility depends crucially on when fertility is assessed.} A second innovation of our study is that we examine the response to \textit{individual} shocks, complementing previous work on the effects of war on entire cohorts or regions. Among the shocks are war captivity and injury, two widespread but hardly studied war experiences.
To our knowledge, the only evidence on the economic impact of these wartime shocks comes from the American Civil War of 1861-65 \citep{Lee2005,Costa2012,Costa2020} and this evidence confirms the importance of these shocks for socioeconomic outcomes later in life.\footnote{More evidence exists on the impact of war captivity on health and mortality, although many studies show only statistical associations; see \cite{myrskyla2023look} for recent causal evidence from the Finnish Civil War.} In addition, we highlight that the consequences of displacement for education and labor market careers interact critically with the age at displacement, adding to a growing literature documenting heterogeneity in the impact of displacement \citep[][see \citeauthor{Becker2019}, \citeyear{Becker2019}, and \citeauthor{Becker2022}, \citeyear{Becker2022}, for recent surveys of forced displacement]{Bauer2013,Bauer2019,Sarvimaeki2022}.

Our study also relates to the literature on human capital and labor supply decisions \citep{BenPorath1967, heckman1976life}. Interest has focused on whether standard models can explain important empirical regularities, such as the shape of age-experience profiles in earnings \citep{Mincer1997}, or long-run trends in schooling and labor supply \citep{Hazan2009,Cervellatietal2013}. In contrast, there exists less evidence on whether such models can explain the response to exogenous \textit{shocks}. A notable exception is the literature on the effect of diseases on human capital investments and development \citep{bleakley2010health, fortson2011mortality, ManuelliYurdagul2021}. In comparison, rather than shifts in life expectancy across time and place, we study the response to realized shocks on the individual level.

\section{Data and Background}\label{sec:data}

\paragraph{German Life History Study.} Our main data source is the GHS, a survey of eight West German birth cohorts born 1919-1971 \citep{Mayer2007}. We draw on retrospective life-cycle information\footnote{Recall error can increase the measurement error of retrospectively collected data, especially for distant events. However, multiple studies have extensively assessed the reliability of retrospective data in the GHS, with quality enhanced through careful data editing  \citep[see][for details]{Mayer2007}. Event histories were checked for completeness, consistency, and plausibility across life domains, with interview recordings reviewed or respondents recontacted when needed. Importantly, our treatment indicators do not rely on the exact recall of war injury, imprisonment, or displacement dates.} from the first two waves \citep{ZA2647,ZA2646,ZA2645}. Both waves constructed nationally representative samples of German citizens living in West Germany at the time of the survey (foreigners were excluded). The first wave (GHS-1, conducted in 1981-83) surveyed  respondents born in 1929-31, 1939-41, and 1949-51. We instead focus on the second wave (GHS-2, conducted in 1985-88), which surveyed individuals born in 1919-21, who were 18-20 years old when the war started. We observe 1,412 respondents in this wave, 559 of which are men (the gender imbalance is a legacy of WWII). Although modest, this main sample is sufficiently large to capture the effects of the life-changing shocks we are studying. Moreover, parts of our analysis can be also implemented in the much larger 1970 census, as discussed below.

Importantly for our purpose, the GHS-2 focused on events related to WWII. It recorded occupational absences and gaps in employment due to military service, captivity, and displacement. Respondents were also asked about illness, accidents, or ailments. We classify bullet and shrapnel wounds, frostbite, amputations, and ``general war injuries'' sustained between 1939-45 as war-related injuries. 

The GHS also contains rich demographic and labor market information. Respondents reported their complete education, employment, and family history. 
For each life domain, respondents were asked about sequences of episodes, including start and end dates and additional information about each spell. For example, in the employment domain, respondents provided the start and end dates of each employment spell in chronological order, as well as information on job title, industry, or hours worked. Once the data were collected for each domain, the information was checked for consistency across life domains.
 
To examine occupational success, we use the Standard International Occupational Scale \citep{Treiman1977}, in which the values for occupational prestige range from 18 (unskilled workers) to 78 (doctors, professors) and are coded as missing for periods of non-employment. The GHS recorded time spent in school, vocational training, and further education, and we measure educational attainment by total years of education (see Online Appendix \ref{sec:Education-Measure} for details). We also observe education and employment outcomes for the respondents' parents. In addition, the GHS-2 recorded pension income, distinguishing between different sources, including pensions from own work  and war victims' pensions (we discuss war compensations below). When analyzing pension incomes, we restrict attention to GHS-2 respondents surveyed in 1987/88 and thus at age 66 or older.\footnote{\label{fn:lvs}The GHS-2 was conducted in two parts. The first part surveyed 407 respondents in 1985/86 using face-to-face interviews; the second surveyed 1005 respondents in 1987/88 using telephone interviews.} 

The GHS recorded the complete residential history of respondents, which we use to define displacement status. As in official population statistics, we define displaced persons as Germans who lived in the eastern territories of the German Reich, Czechoslovakia, or Eastern Europe on 1 September 1939.\footnote{For individuals born in 1939-41 but after September 1939, we use their residence at birth.} 
We classify all other individuals as non-displaced, except for Germans living in the future Soviet occupation zone in 1939, whom we exclude from this part of the analysis as they were positively selected in education \citep{Becker2020a}. Overall, 7.9 million displaced persons (so-called expellees) lived in West Germany in 1950 \citep{StatistischesBundesamt1955}. Most arrived in 1945-46, primarily from the eastern territories of the German Reich, which Germany ceded in 1919 and 1945 (see Online Appendix Figure \ref{fig:TerritorialLosses} for an overview of Germany's territorial losses). 

\paragraph{1970 census.} Two limitations of the GHS are its small sample size and the focus on specific birth cohorts, limiting its usefulness in examining how individual war experiences vary across cohorts. For this type of analysis, we therefore use a second data set, the West German population and occupation census of 1970. The data comprise a 10\% random sample of the population, almost 6.2 million individuals \citep{RFSOSO}. The census contains information on an individual's residence on 1 September 1939 (or the father's residence for individuals born after that date). As in the GHS, we drop migrants from the Soviet occupation zone and define displaced persons as Germans who lived in Germany's former eastern territories, Czechoslovakia, or Eastern Europe in 1939. Unfortunately, we cannot identify war wounded or POWs in the census. 

The census provides information only on socioeconomic attainment in 1970. Two outcome variables are nevertheless of interest for our analysis: educational attainment and the year of exit from employment. We measure years of education by adding the years spent in vocational training and university to the time required to attain the highest school degree. The census also asked respondents who were not employed in 1970 when they left their last job. For older individuals over the statutory retirement age of 65, we can be reasonably confident that the year indicates the end of their labor market career. 

\paragraph{Exposure to wartime shocks.} The shocks we consider--war injuries, captivity, and displacement--were common (see Table \ref{tab:prevalenceshocks}). In the GHS sample of men born in 1919-21, 29.9\% suffered from war injuries, most commonly bullet wounds. Our empirical analysis compares injured and non-injured soldiers over their life cycles. Among men born in 1919-21, three-fourths were POWs, but the length of imprisonment varied considerably, with a mean of 16.5 and a standard deviation of 20.3 months. Our analysis compares the life-cycle profiles of soldiers who were POWs for more than six months to those who were not imprisoned (our results remain similar when including POWs imprisoned for fewer than six months). The treated make up 47.4\% of respondents. Finally, more than one in five individuals in our GHS sample were displaced (the share is similar in the 1970 census). We compare them to their non-displaced peers (after excluding GDR refugees). We discuss potential concerns related to selection and multiple treatments in the next section.

	\begin{table}[t]
		\centering
		\begin{threeparttable}
			\caption{Exposure to individual war shocks} \centering
			\label{tab:prevalenceshocks}
			\begin{footnotesize}
				\begin{tabular}[c]{lcccccccccc}
					\toprule	
					&  \multicolumn{2}{c}{War injuries}  && \multicolumn{3}{c}{War captivity} && \multicolumn{3}{c}{Displacement}  \\ 
					&  \multicolumn{2}{c}{(men)}  && \multicolumn{3}{c}{(men)} && \multicolumn{3}{c}{(men and women)}  \\ 					
					&  \multicolumn{2}{c}{1919-21}  && \multicolumn{3}{c}{1919-21} && 1919-21 & 1929-31 & 1939-41  \\ 
					\cline{2-3} \cline{5-7} \cline{9-11}
					&  &Bullet &&  & $>$6 &  Length &&  \multicolumn{3}{c}{} \\   
					& Share  &wounds && Share &   months & (months)&& Share & Share & Share \\ 
					& (1) & (2) && (3) & (4) & (5) && (6)  & (7) & (8)  \\	\hline
					Mean		& 0.299 &0.206 &&  0.755 & 0.474  & 16.547 && 0.227 & 0.182 & 0.187 \\
					Std. deviation		& (0.458) & (0.405) && (0.431)& (0.500) & (20.256) && (0.419) & (0.386) & (0.390)  \\
				Observations 	& 559      & 559  &    & 559      & 559	&   559   &  & 1,278     &  661   &  673 \\                   					
					\bottomrule		
				\end{tabular}
			\end{footnotesize}
			\begin{tablenotes}[flushleft]
				\item \footnotesize{\emph{Notes}: Statistics for battlefield injuries and war captivity are based on men born 1919-21. Statistics for displacement are based on men and women, excluding GDR refugees.}
                \item \footnotesize{\emph{Source}: \cite{ZA2647, ZA2646, ZA2645}}
			\end{tablenotes}
		\end{threeparttable}
	\end{table}

\paragraph{Compensation programs.}  Not least because of the large number of people affected, the fate of the war-damaged and the displaced was one of the most pressing social problems in the postwar period. The importance of this problem is reflected in the fact that a newly created ministry, the Federal Ministry for Displaced Persons, Refugees and War-Affected Persons, was charged with solving it. Policymakers developed two critical programs \citep{Wiegand1995} to compensate war victims and facilitate their integration into the labor market, the war victims' provision (\textit{Kriegsopferversorgung}) and the equalization of burden (\textit{Lastenausgleich}). 

The financially most significant part of the war victims' provision was the war victim's pension (\textit{Kriegsopferrente}). The pension was paid to persons who suffered serious health damage as a result of military or military-like service in connection with the war (e.g. damage due to direct warfare, captivity, or internment abroad). The war victims' provision also paid for measures suitable for improving the earning capacity of war-disabled persons, such as advanced training and vocational (re)training, and for curative treatments to restore their health. Overall, the war victims' compensation consumed a significant share of the government budget, especially in the immediate post-war period when it accounted for up to 16\% of total social expenditure in Germany (see Online Appendix \ref{sec:compensation} for details).   
	
The war victims' pension had initially two main components. First, a basic pension (\textit{Grundrente}) was paid to all war-disabled persons whose civilian earning capacity was reduced by at least 25\%. Second, an additional equalization pension (\textit{Ausgleichsrente}) was paid to severely disabled persons whose earning capacity was reduced by 50\%, if they were no longer able to work (or only to a limited extent). Unlike basic assistance, the equalization pension was means-tested. The exact amount of the basic and equalization pensions depended on the severity of the war-related health damage and was regularly adjusted. Appendix \ref{sec:compensation} discusses the evolution of pension levels in detail.

In 1956, about 1.25 million war-damaged received a basic pension and another 0.25 million received both a basic and an equalization pension \citep{PIB1957}.\footnote{In addition to the war-damaged, 1.17 million widows and 1.09 million orphans received war victims' pension. See \cite{Braun2024} for an analysis of the employment biographies of war widows in West Germany after 1945.} Benefits were gradually expanded, and from 1960 severely disabled persons received an additional damage compensation (\textit{Schadensausgleich}) if their income was below of what they would have earned without the war damage. Importantly, the war victim's pension was not tied to any age thresholds. 

The second program, equalization of burdens, partially compensated for the loss of wartime property, thereby distributing the burdens of war more evenly throughout society. Those whose property had been spared by the war were to compensate those who had suffered war damage. Displaced persons could also apply for grants to start businesses and for public assistance in finding housing. However, the equalization of burdens had only limited success in restoring the occupational status of the displaced and the prewar distribution of wealth \citep{Falck2012,Wiegand1995}.

\begin{figure}[!tb]
	\caption{Life-cycle profile for the male cohort born in 1919-21}\label{fig:lc_profiles_1919_21}
	\centering
	\begin{threeparttable}
		\includegraphics[width=0.75\textwidth]{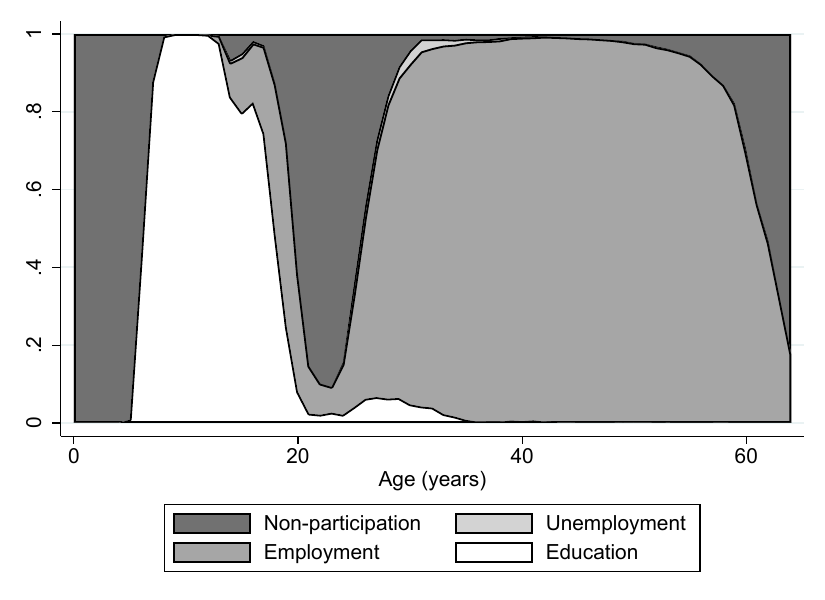}\\
		\begin{tablenotes}[flushleft]
			\item \footnotesize{\emph{Notes}: The figure shows the share of individuals in four mutually exclusive states: non-participation, unemployment, employment, and education (including school, vocational training, and further education).}
            \item \footnotesize{\emph{Source}: \cite{ZA2647, ZA2646}}
		\end{tablenotes}
	\end{threeparttable}
\end{figure}

\paragraph{Cohort effects.} Figure \ref{fig:lc_profiles_1919_21} plots the life-cycle profile for males born in 1919-21, distinguishing four states: non-participation, unemployment, employment, and education. The cohort was 18-20 years old when the war started, and all males were conscripted.\footnote{The 1919 cohort was conscripted on 26 August 1939, the 1920 cohort on 1 October 1940, and the 1921 cohort on 1 February 1941 \citep{Kroener1988,Kroener1999}.} War service thus hit the cohort when, without compulsory military service, as was the case during the Weimar Republic, 
it would have entered the labor market, severely interrupting the transition from education to work \citep{Brueckner1987}. The figure shows that the cohort's labor force participation rate plummeted abruptly around age 20 and gradually recovered in the late 20s. War captivity reduced the cohort's labor force participation long after the war ended.

Online Appendix Figure \ref{fig:lc_profiles_191951} compares the life-cycle profile of the 1919-21 cohort with later-born cohorts covered by the GHS-1, illustrating that their transition from education to work was dramatically different.\footnote{See also \cite{Mayer1988} for a cohort perspective of the impact of WWII on German survivors.} Men born in 1919-21 spent, on average, just 156 months in the labor market by age 37, more than 60 months less than males born in 1929-31 or 1939-41. The latter cohorts were either not (1929-31) or only partially (1939-41) conscripted after West Germany rearmed in 1955.\footnote{Compulsory military service, initially for 12 months, was reintroduced in 1956, with the first draft into the \textit{Bundeswehr} in 1957 for men born after June 1937. In the GHS, only 22\% of men born 1939-41 report having completed basic military service (\textit{Grundwehrdienst}) or alternative service (\textit{Ersatzdienst}).} In contrast, women born in 1919-21 had higher participation rates than later cohorts (see Online Appendix \ref{sec:cohorteffects}). Our analysis abstracts from cohort effects and instead identifies within-cohort differences by war experience. 

Figure \ref{fig:lc_profiles_1919_21} also shows that most men born in 1919-21 left employment well before the statutory retirement age of 65. Several early retirement paths existed at the time (see the appendix for details). Individuals with more than 35 years of contributions could retire at 63, while those who were long-term unemployed or severely disabled could retire at 60. The pension system also provided disability benefits for workers of any age, subject to medical indication and a minimum insurance period of at least five years.

\paragraph{Macroeconomic conditions.} Macroeconomic conditions were favorable for the integration of former soldiers into the labor market. West Germany recovered swiftly from WWII \citep{Eichengreen2009}, and real GDP per capita nearly tripled between 1950 and 1970. The mass inflow of displaced persons increased unemployment initially \citep{Braun2014}, but provided a valuable pool of labor for the booming economy in the late 1950s and 1960s. Unemployment fell steadily, from 10.4\% in 1950 to 1.3\% in 1960, and remained very low until the mid-1970s. These favorable conditions likely explain why, despite their late entry into the labor force, men in the 1919-21 cohort were rarely unemployed after age 35. As discussed, policy measures also improved the earning capacity of the war-disabled, by covering the costs of vocational training and curative treatments.

\section{Life-Cycle Effects of Individual War-Time Shocks}\label{sec:results1} 

This section considers the impact of battlefield injuries, war captivity, and displacement on labor market outcomes of men born 1919-21. For each shock, we present three sets of results. First, we report cross-sectional summary measures of labor market success, based on regressions such as 
\begin{equation}\label{eq:1}
y_i = \alpha + \beta shock_i + \gamma x_i + \varepsilon_i,
\end{equation}
where $y_i$ is an outcome of interest for individual $i$, $shock_i$ is one of the three shocks listed in Table \ref{tab:prevalenceshocks}, and $x_i$ is a set of control variables as defined below. In robustness tests, we also include all three shocks jointly to account for the fact that individuals might be exposed to more than one shock. This has little effect on our estimates, as the shocks are hardly correlated with each other (as discussed below).

We restrict the sample to WWII veterans when examining war injuries and imprisonment. In these cases, our control group consists of other soldiers who also served in WWII, but who did not suffer a war injury or become a POW. Thus, we are not examining the effect of WWII service per se but the effect of specific war experiences conditional on military service. This focus is of particular interest for countries such as Austria or Germany, where public policy has targeted war-disabled veterans rather than veterans in general. As for the effect of war service in general, \cite{Maas1999} show for German men born between 1887 and 1922 that the duration of military service had a negative effect on men's occupational careers in the short run, but quickly vanished as the German economy recovered.

Second, we show full life-cycle plots for employment and occupational prestige, based on pooled regressions that interact the shock with age: 
\begin{equation}\label{eq:2}
y_{it} = \sum_t \alpha_t I(Age=t) + \sum_t \beta_t I(Age=t) shock_i + \sum_t  \gamma_t I(Age=t) x_i + \varepsilon_{it},
\end{equation}
where $y_{it}$ is the employment status or occupational prestige of individual $i$ at age $t$ (age in years). Third, we ask whether individuals' responses are consistent with the predictions from standard life-cycle theory of human capital and labor supply decisions. For readability, most of this theoretical discussion is relegated to the online appendix, with only the main results summarized in the main text. 

\paragraph{Identification.}  A concern for identification is that war shocks may correlate with individual characteristics affecting labor market supply or prospects after the war ended. To address this concern, we show that battlefield injuries and imprisonment were unrelated to own and family characteristics, in line with historical studies.

Enlistment was near-universal for the 1919-21 cohort \citep{Overmans1999}, with 95\% of men in our sample fighting in the war. Neither the duration nor the intensity of their exposure to the war varied much with family or own characteristics. While qualified workers in the arms industry were initially spared from military service \citep{Mueller2016}, the 1919-21 cohort was still too young to fill such positions when WWII began. Indeed, their year of war entry differs little in our sample, as the 1919 and 1920 cohorts formed the backbone of Hitler's Wehrmacht at the beginning of the war. Highly educated individuals were as likely to be placed in combat roles as their less educated peers, and social background had little to do with the assignment of individuals to subunits--and thus their risk of injury or imprisonment.\footnote{For example, \cite[p. 686]{Fritz1996} notes that ``\textit{Unlike the American army, which until 1944 shunted its most educated men into specialized roles, the Wehrmacht from the start deployed a remarkably high percentage of its manpower as combat troops}''. This was especially true of the young, on whom our analysis focuses. Similarly, \cite{Rass2003} finds no association between social origin and allocation across subunits in his study of a German infantry division. The exception was the intelligence unit, which recruited predominantly from higher socio-economic classes, but accounted for only 3\% of the division's personnel.}   

The 1919-21 cohort was also too young to reach the middle and higher officer ranks, who had been less likely to die in past wars than regular soldiers. In our data, only 5\% have reached lieutenant rank. Even the lower officer ranks of captains and majors tended to be significantly older than the 1919-21 cohort, averaging 33.5 and 26.5 years of age in 1942, respectively \citep{Foerster2009}. Moreover, unlike in previous wars, there was a high casualty rate among officers in WWII \citep{Mueller2016}. 
For example, the proportion of officers among all missing during the deadliest defeat in German military history, the collapse of the \textit{Heeresgruppe Mitte} on the Eastern Front in 1944, corresponded exactly to their proportion in two combat divisions studied by \cite{Hartmann2010}.\footnote{``\textit{This suggests that the 1944 catastrophe must have affected the [divisions] across the board. All became their victims, regardless of rank, function, or place of operation}'' \citep[p. 220, own translation]{Hartmann2010}.} 

Instead, the likelihood of injury and duration of imprisonment depended primarily on which part of the front the soldiers were fighting,\footnote{While all POWs in Western Allied custody were released by the end of 1948, the last POW from Soviet captivity did not return until 1956. In our sample, those serving (also) at the Eastern front were more than twice as likely to suffer from bullet wounds than those serving only at other fronts.} over which they had little control. Notably, the soldiers' region of deployment did not depend on their regions of origin \citep{Overmans1999}. And while the very young and old were more likely to be released early from captivity \citep{Overmans2000}, our analysis conditions on age and focuses on a birth cohort that was already in their mid-20s by the end of the war.
 
As for the displacement effect, \cite{Bauer2013} show that the differences between displaced and non-displaced Germans were small before the war, not least in education. The only major differences were in the proportions employed in agriculture and industry, which can be attributed to the more agrarian structure of the eastern territories. However, the 1919-21 cohort had little labor market experience when the war began, so these structural differences had little impact on their work experience. Importantly, virtually all Germans living east of the postwar German-Polish border were displaced, minimizing problems of selection.

We provide evidence in support of these arguments in panel (a) of Table \ref{tab:exoshocks}, regressing indicators for war service and each shock on prewar characteristics of the respondents (birth year, siblings, years of schooling, and an indicator for ill health in childhood) and their parents (years of education and the father's occupational score). Individuals who were sick in childhood were less likely to serve in the war, but war service is uncorrelated with socio-economic characteristics. This lack of selectivity is in line with the observation that enlistment was nearly universal for the cohorts that we study. 
It is also consistent with earlier findings in \cite{Maas1999}, who find no evidence of an effect of socio-economic characteristics on the likelihood and length of WWII service in a sample of German men born 1887-1922.

Conditional on serving, prewar characteristics do not predict war injuries, captivity, or displacement, explaining less than 2\% of their variation. As an exception, individuals who report periods of poor health before age 18 tend to be less exposed to the war shocks we study. However, few individuals report poor health before age 18 (see Online Appendix Figure \ref{fig:injury_illness}), and our estimates change little when controlling for pre-war health.\footnote{Since about one-third of POWs in Soviet custody died, one might expect survivors to be positively selected on health. However, the mortality rate of German POWs varied greatly with the place and time of internment, limiting selectivity with respect to individual characteristics and possibly explaining the lack of correlation with pre-war socio-economic status (Table \ref{tab:exoshocks}). Returning POWs reported similar or \textit{worse} health at later ages (Online Appendix Figure \ref{fig:injury_illness}).} The patterns remain similar when estimating separate bivariate regressions for each characteristic (Online Appendix Table \ref{tab:exoshocksbi}). Our baseline analysis nevertheless controls for birth year indicators, parental education, number of siblings (which correlates with socio-economic background), and time of war entry. In an extended specification, we additionally control for own years of secondary schooling, poor health before age 18, and the respective other war shocks. The model fit increases as controls are added while our coefficients of interest remain stable, making it unlikely that omitted variables drive our results.

\begin{table}[!t]
	\begin{center}
		\begin{threeparttable}
			\caption{Exogeneity of war service and war shocks} \centering
			\label{tab:exoshocks}
			\begin{footnotesize}	
				\begin{tabular}{lccccc}
					\toprule 
					& & War service & War injury & War captivity & Displaced \\
					& & (0/1) & (0/1) & (0/1) & (0/1) \\
					& mean   & men & men & men & men \& \\
					&  (std. dev.) & only  & only &  only & women \\
					&  & (1) &(2) &(3)&(4)\tabularnewline
					\midrule
					\multicolumn{4}{l}{\textbf{(a) Pre-war characteristics}} \tabularnewline
					Father's years of schooling & 8.68 & 0.004 & -0.002 & 0.004 & -0.006 \\
        					& (2.07) & (0.004) & (0.015) & (0.014) & (0.009) \\
        					Mother's years of schooling & 8.24 & -0.013 & -0.019 & 0.011 & 0.008 \\
       				 	& (0.95) & (0.016) & (0.034) & (0.025) & (0.017) \\
       					Father's occupational score & 41.0 & -0.001 & 0.003 & -0.004 & 0.000 \\
        					& (10.9) & (0.001) & (0.003) & (0.002) & (0.001) \\
        					Birth year & 1920.1 & 0.014 & -0.028 & 0.019 & 0.003 \\
        					& (0.79) & (0.013) & (0.029) & (0.025) & (0.016) \\
        					\# siblings & 2.86 & -0.004 & 0.005 & 0.011 & 0.004 \\
       					& (2.44) & (0.004) & (0.010) & (0.008) & (0.005) \\
        					Years of schooling & 8.77 & 0.006 & -0.029* & 0.019 & 0.019 \\
        					& (1.42) & (0.007) & (0.017) & (0.015) & (0.012) \\
        					Poor health before age 18 & 0.08 & -0.234*** & -0.088 & -0.087 & -0.077* \\
        					& (0.27) & (0.074) & (0.086) & (0.095) & (0.040) \\
        					Female & 0.60 &  &  &  & 0.003 \\
        					& (0.49) &  &  &  & (0.026) \\
        					\midrule
					Observations & 1,412 & 492 & 465 & 465 & 1,054 \\
        					R2 &  & 0.084 & 0.016 & 0.014 & 0.006 \\
					\midrule			
					\multicolumn{4}{l}{\textbf{(b) Position in the military}} \tabularnewline				
					Branch: Navy (cf: Army) 				& 0.08 	& --& -0.464*** & 0.244*** & 0.040 \tabularnewline
					& (0.27) 	& &  (0.131) & (0.052) & (0.148) \tabularnewline
					Branch: Air force 					& 0.30 	& --& -0.410*** & 0.022 & -0.082 \tabularnewline
					& (0.46) 	& & (0.081) & (0.072) & (0.077) \tabularnewline
					Rank ($\ge$ \textit{Unteroffizier}) 		& 0.49 	& --& 0.052  & -0.016 & 0.028 \tabularnewline
					& (0.50) 	& & (0.077) & (0.062) & (0.066) \tabularnewline
					Volunteer (cf: drafted) 				& 0.33 	& --& 0.031 & -0.102 & 0.132* \tabularnewline
					& (0.47) 	& & (0.078) & (0.077) & (0.078) \tabularnewline
					Specialized training 					& 0.67 	& --& 0.041 & 0.012 & -0.076 \tabularnewline
					& (0.47) 	& & (0.088) & (0.071) & (0.079) \tabularnewline \hline
					Observations 	& 178 	& --& 175 & 177 & 167 \tabularnewline
					R2 	&  		&   & 0.156 & 0.035 & 0.035 \tabularnewline
					\bottomrule
				\end{tabular}
			\end{footnotesize}
			\begin{tablenotes}[flushleft] \item \footnotesize{\emph{Notes}: The table reports coefficient estimates from the indicated war-related shock on a set of pre-war individual and parental characteristics (panel (a)) or indicators of their position within the military during the war (panel (b)) for birth cohorts 1919-21. Estimates for war injuries and captivity are conditional on war service. Robust standard errors in parentheses. ***, **, and * denote statistical significance at the 1\%, 5\%, and 10\% level, respectively.}
            \item \footnotesize{\emph{Source}: \cite{ZA2647, ZA2646}}
			\end{tablenotes}
		\end{threeparttable}
	\end{center}
\end{table}

For a subset of our data, we also observe the individuals' military rank, training, and branch, and whether they volunteered to serve. Rank or specialized military training (beyond basic combat training) are uncorrelated with any of our war shocks (Table \ref{tab:exoshocks}, panel (b)). As expected, army soldiers were much more likely to sustain war injuries than those serving in the navy or the air force, and sailors were imprisoned more often than soldiers of the other two branches. However, individual or parental characteristics do not predict military branch or voluntary service, and do therefore not predict war injuries or captivity either (Table \ref{tab:exoshocks}, panel (a)). Overall, we find no evidence of significant selection into treatment. Furthermore, the specific lifecycle patterns we document--where treatment effects are concentrated at certain ages--would be difficult to attribute to selection alone.

\paragraph{Multiple shocks.}  One other concern is that soldiers can be subject to multiple war-related shocks. If the different shocks were correlated, we would risk capturing the consequences of multiple war experiences rather than a clearly defined shock (omitted variable bias formula). Online Appendix Table \ref{tab:corrshocks} shows that this is not a concern in our setting. Displacement is not correlated with either imprisonment or wartime injuries. Imprisonment and wartime injuries show a slight negative correlation, presumably because wartime injuries reduced subsequent combat time and thus the risk of imprisonment.\footnote{More generally, the shocks we study may have altered subsequent combat exposure. For example, capture by the Western Allies or hospitalization might have reduced the risk of further traumatic combat experiences, which in turn would have had health and possibly economic costs \citep[e.g.][]{Cesuretal2013,Edwards2015}. While this does not undermine the causal interpretation of our estimates (as the reduced combat exposure results from the shocks), it must be kept in mind when interpreting them.} However, the correlation is small ($\rho=-0.11$), and we show in additional checks that our estimates change little when adding the other respective shocks as controls. 

Nevertheless, the different war shocks might interact with each other. For example, war injuries might have a different effect on displaced than on non-displaced soldiers. In Online Appendix Section \ref{sec:multishocks}, we show that even if such interaction effects exist, our approach captures an appropriately weighted average effect. For example, our estimate for war injuries reflects the weighted average of their effects on displaced and non-displaced soldiers, with weights equal to their respective population shares. Online Appendix Section \ref{sec:multishocks} also presents separate estimates for combinations of war-related shocks. However, because cell sizes are small for some combinations of shocks, our main analysis focuses on the average effect of each shock, as identified by equations (\ref{eq:1}) and (\ref{eq:2}).

\paragraph{Selective mortality.} One might worry that veterans with serious injuries were more likely to die before the 1980s and thus underrepresented in the retrospectively collected GHS. Two observations suggest that such selective mortality is not an issue for our analysis. First, official mortality tables for the postwar period do not show an unusually high risk for males born around 1920 \citep{StatistischesBundesamt2006}. For example, the 1920 male cohort had a remaining (post-war) life expectancy of 47.0 years at age 26, only about 1.6 year less than the non-serving cohort of 1930. This increase in life expectancy between cohorts is not unusually large (life expectancy at age 26 increased by two years from the 1930 to the 1940 cohort). Second, the share of injured veterans in our data is broadly in line with existing estimates. \cite{Mueller2016} states that about 5.2 million German soldiers were injured in the war (out of 18.2 million serving, corresponding to a share of 28.5\% compared to 29.9\% in our sample). Even with selective mortality we would still estimate a lower bound on the true (negative) effect of war injuries on labor market outcomes.\footnote{The GHS allows us to distinguish severely from less severely injured veterans, and the labor market consequences are more pronounced for the former. If the severely injured faced a higher mortality risk, we would, therefore, underestimate the true effect of war injuries. Note also that selective mortality could not affect the direction of our results unless mortality were substantially higher \textit{and} at-risk veterans showed the opposite labor market effects than less severely injured veterans; our data support neither of these assumptions.}

\paragraph{Battlefield injuries.} Panel (a) of Table \ref{tab:cs191921} summarizes how battlefield injuries affected the labor market careers of WWII veterans born 1919-21. The estimates are based on equation (\ref{eq:1}), estimated separately for each outcome and shock (the estimates change little when we include all shocks jointly, as shown below). Two main findings emerge. First, injured soldiers were less likely to be employed at older ages than non-injured veterans, although they achieved similar employment rates at earlier ages (see columns (2) and (3)). War injuries reduced employment by nearly one year between ages 56-65, by accelerating the transition from work to retirement. Second, war injuries reduced monthly work pensions by DM 234 or almost 10\% compared to the control mean (column (5)). However, the higher pension payments as war victims almost compensated for this loss (column (6)). Battlefield injuries have no sizeable effects on educational investments (column (1)) or on nonpension income in old age (column (7)). %

\begin{table}[t!]
	\hskip -7pt
	\begin{threeparttable}
		\caption{The effect of war experiences on labor market outcomes, men born 1919-21} \centering
		\label{tab:cs191921}
		\begin{footnotesize}
			\begin{tabular}{lccccccc}
				\toprule
				&Educational& \multicolumn{2}{c}{Years in employment} & Occup.  &  \multicolumn{3}{c}{Old age income from}  \\ \cline{3-4} \cline{6-8}
				&attainment&  \hspace{0.4cm}age\hspace{0.4cm}  & \hspace{0.3cm}age\hspace{0.3cm} & prestige  & work  & war victim & nonpension\\
				&(years)& 20-55 & 56-65 & (maximum) & pension  & pension & sources \\
				& (1)	& (2)	& (3)  & (4) & (5) & (6) & (7)  \\
				\midrule
				\multicolumn{4}{l}{\textbf{(a) War injury (0/1)}}  & & & & \\
				  &      -0.263   &       0.060   &      -0.908***&      -0.086   &    -233.91*  &     180.63***&     -14.98   \\
				  &  (0.199)   &  (0.320)   &  (0.290)   &  (0.987)   &(122.80)   & (48.76)   &(131.74)   \\
				  &     [11.06]   &      [28.68]   &     [6.03]   &      [46.90]   &    [2390.73]   &      [20.31]   &     [375.97]   \\
				Observations	& 465 & 465 	  & 465 	 & 465 		& 282 		& 282 		& 297 \\
				\bottomrule
				\multicolumn{4}{l}{\textbf{(b) War captivity ($>$ 6 months)}}  & & & & \\
				    &      -0.262   &      -2.266***&       0.446   &      -2.879** &     -11.71   &    -121.35***&     -62.98   \\
				    &     (0.224)   &     (0.332)   &     (0.332)   &     (1.201)   &   (129.94)   &    (42.21)   &   (137.43)   \\
				    &      [10.98]   &      [29.75]   &       [5.29]   &      [48.36]   &    [2301.65]   &     [146.09]   &     [377.35]   \\
				Observations	& 331  & 331 	& 331  & 296   & 203 		& 203 		& 216 \\
				\bottomrule
				\multicolumn{4}{l}{\textbf{(c) Displacement (0/1)}}  & & & & \\
				  &      -0.150   &      -0.638*  &       0.121   &      -2.424** &     -80.08   &      19.74   &    -310.05***\\
				  &     (0.233)   &     (0.367)   &     (0.314)   &     (1.015)   &   (144.09)   &    (41.30)   &    (94.47)   \\
				  &      [10.93]   &      [28.82]   &       [5.69]   &      [47.04]   &    [2376.11]   &     [59.57]   &   [459.16]   \\
				Observations 	& 427 		& 427  		& 427 & 427  		& 254 		& 254 		& 269  \\
				\bottomrule   
			\end{tabular}%
		\end{footnotesize}
		\begin{tablenotes}[flushleft]
			\item \footnotesize{\emph{Notes}: Estimates of the effect of battlefield and other war-related injuries (panel (a)), war captivity (panel (b)), and displacement (panel (c)) on various outcome variables (shown in the table header). Each estimate is from a separate regression. The sample consists of males born 1919-21. Columns (5) to (7) restrict the sample to the second part of GHS-2 conducted in 1987/88 where we observe individuals up to age 65-69 (see Footnote \ref{fn:lvs}). All regressions control for the birth year (indicators), years of schooling of father and mother, number of siblings, and time of entry into the war. Robust standard errors in parentheses, unconditional means for the unaffected control group are in square brackets. ***, **, and * denote statistical significance at the 1\%, 5\%, and 10\% level.}
            \item \footnotesize{\emph{Source}: \cite{ZA2647, ZA2646}}
		\end{tablenotes}
	\end{threeparttable}
\end{table}

\begin{figure}[!tbp]
	\caption{Life-cycle effects of war experiences, men born 1919-21}\label{fig:lfeffects_191921}
	\centering
	\begin{threeparttable}
		\begin{center}
			\textit{\begin{center}
					War injury\\ \smallskip
			\end{center}}
			\subfloat[Employment]{\includegraphics[clip, trim=0.3cm 0.7cm 0.3cm 0.7cm, width=0.48\textwidth]{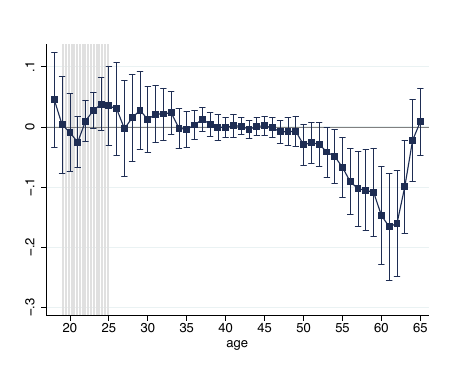}\label{fig:injury_lm2}}
			\hskip 7pt
			\subfloat[Occupational Prestige]{\includegraphics[clip, trim=0.3cm 0.7cm 0.3cm 0.7cm, width=0.48\textwidth]{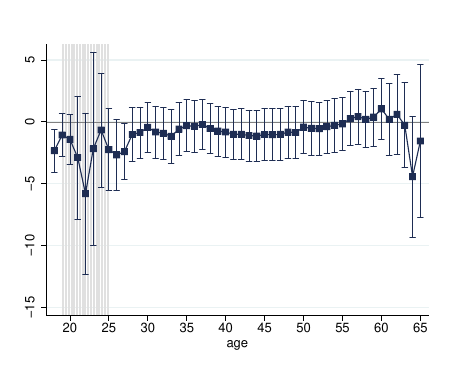}\label{fig:injury_occ}}\\
			\textit{\begin{center}
					War captivity\\ \smallskip
			\end{center}}
			\subfloat[Employment]{\includegraphics[clip, trim=0.3cm 0.7cm 0.3cm 0.7cm, width=0.48\textwidth]{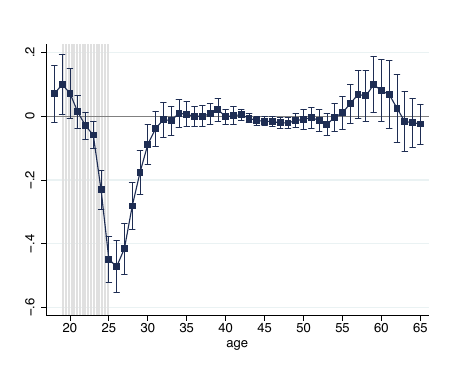}\label{fig:pow_lm2}}
			\hskip 7pt
			\subfloat[Occupational prestige]{\includegraphics[clip, trim=0.3cm 0.7cm 0.3cm 0.7cm, width=0.48\textwidth]{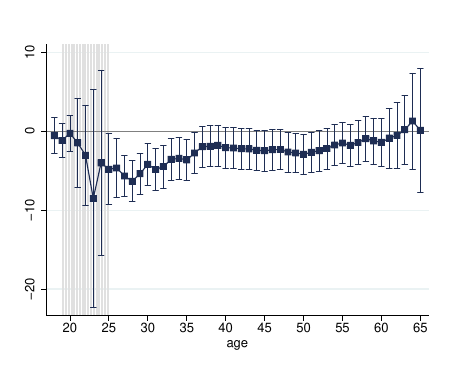}\label{fig:pow_occ}}\\
						\textit{\begin{center}
					Displacement\\ \smallskip
			\end{center}}
			\subfloat[Employment]{\includegraphics[clip, trim=0.3cm 0.7cm 0.3cm 0.7cm, width=0.48\textwidth]{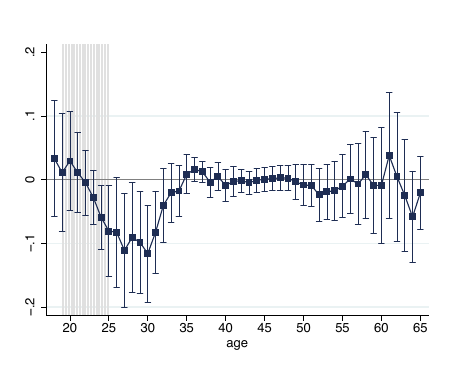}\label{fig:expellee_participation}}
			\hskip 7pt
			\subfloat[Occupational prestige]{\includegraphics[clip, trim=0.3cm 0.7cm 0.3cm 0.7cm, width=0.48\textwidth]{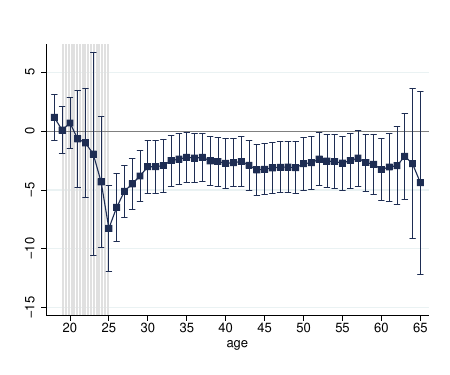}\label{fig:expellee_prestige}}
		\end{center}
		\begin{tablenotes}[flushleft]
			\item \footnotesize{\emph{Notes}: Effect of battlefield and other war-related injuries (panels (a) and (b)), war captivity (panels (c) and (d)), and displacement (panels (e) and (f)) on employment (left panels) and occupational prestige (conditional on employment, right panels) over the life cycle. Estimates are from a pooled OLS regression, interacting the regressor of interest and birth year (indicators) with a full set of age indicators. The sample consists of males born 1919-21. Point estimates are marked by a dot. The vertical bands indicate the 95\% confidence interval of each estimate. The shaded area indicates the duration of WWII.}
            \item \footnotesize{\emph{Source}: \cite{ZA2647, ZA2646}}
		\end{tablenotes}
	\end{threeparttable}
\end{figure}

The life-cycle graph in panel (a) of Figure \ref{fig:lfeffects_191921} confirms that the adverse employment effect of war injuries arose only late in life. Employment probabilities of men with and without injuries were surprisingly similar at young and middle age\footnote{While the result might combine a negative treatment effect with a positive selection effect, the latter must be small. To see this, write the observed difference in employment status by war injury status ($inj$) as the sum of the average treatment effect on the treated and the selection bias:  $E[y_{mid} \mid inj=1]-E[y_{mid} \mid inj=0] = \left(E[y_{1,mid} \mid inj=1]-E[y_{0,mid} \mid inj=1]\right)+\left(E[y_{0,mid} \mid inj=1]-E[y_{0,mid} \mid inj=0]\right)$. Here, $y_{mid}$ is an indicator for (observed) employment at mid-age, and $y_{1,mid}$ and $y_{0,mid}$ are the potential outcomes with and without having suffered a war injury. Note that $E[y_{0,mid} \mid inj=1]$ can be at most one (the employment rate cannot exceed 100\%). In comparison, between age 35 and 50 the sample estimate (mean) of $E[y_{0,mid} \mid inj=0]$ is 0.984, such that the selection bias is at most 0.016.} but started diverging in the early 50s. The gap then widened steadily until age 61, reaching 17 pp, before shrinking again as injured and non-injured veterans retired. On average, the employment probability of injured veterans was 8.5 pp lower between ages 56 and 65 than that of non-injured peers (relative to a baseline probability of 60.9\%).\footnote{When focusing only on severe injuries, such as amputations, we still find no effect on employment at mid age. Yet, severely injured veterans retired even earlier.} Perhaps surprisingly, we do not find a sustained effect of wartime injuries on occupational success conditional on employment (see Figure \ref{fig:lfeffects_191921}b).
   
As detailed in Online Appendix \ref{sec:theory}, these life-cycle patterns in injured veterans' employment are in line with a standard Ben-Porath type of model with endogenous retirement decisions \citep{Hazan2009}. The model describes how individuals decide to invest in human capital based on their expected returns, and how they time their entry into retirement given its impact on earnings and thus consumption. Specifically, the key equilibrium conditions equate the marginal cost of human capital investment with its marginal benefit over the working life, and the disutility of working at retirement age with the marginal utility of continuing to work (in terms of consumption). In this framework, a natural assumption is that war injuries increase the disutility of work, which tends to decrease employment. But since the disutility of work also increases with age, an upward shift in this disutility tends not to affect early or mid-career employment; instead, it accelerates entry into retirement (see Online Appendix \ref{sec:theory_injuries} and Figure \ref{fig:theory}b).\footnote{As shown in Online Appendix \ref{sec:theory_injuries}, war injuries and the implied reduction in the retirement age also reduce incentives to invest into education. However, the cohort born 1919-21 entered the military around age 20, so many had already completed their educational investments before enlistment.} Standard theory therefore helps to understand why the labor market consequences of a sudden health shock at young age manifest themselves only at older age, decades after the war ended. 

Two factors might amplify this decline in employment at older ages. First, by mitigating the implied income loss, compensations and pensions for war victims increase the incentives for early retirement further (Online Appendix \ref{sec:theory_injuries}).\footnote{See also \cite{Autor2016} who show that a disability compensation program targeting veterans of the Vietnam war greatly reduced their labor supply. The authors note that the response may have been particularly large as the program affected a near-elderly population in diminished health.} Second, health might deteriorate more rapidly over age for wounded than non-wounded veterans (e.g., \citealt{stewart2015retrospective}); Online Appendix Figure \ref{fig:injury_illness} shows that this is the case in our sample. The disutility of work may thus not only shift upwards, but also become steeper over age.\footnote{However, the life-cycle model predicts the labor market consequences of war injuries to be concentrated at older age irregardless of such dynamic health effects. Indeed, controlling for health trajectories, we still find that the war-injured have lower employment in old age than the control group.} 

We also reiterate that disability benefits were available to workers of any age in Germany, subject to medical indication (see the appendix). Disability benefits were an important route to retirement, and their availability likely influenced the precise magnitude and timing of the impact of war injury on early retirement. At the same time, our results show that the large reduction in the retirement age in the 1970s and early 1980s cannot be fully understood without taking into account the fact that at that time cohorts were retiring that had almost entirely served in WWII.

In sum, our findings underscore the importance of a life-cycle perspective in designing policies to alleviate the economic hardship of injured veterans. War injuries can have surprisingly little consequences on labor market careers in the first decades after the war, but then greatly decrease employment rates when veterans reach older ages. As these findings can be rationalized by standard life-cycle theory, we might expect similar patterns following recent or current violent conflicts and wars. 

\paragraph{Prisoners of war.} Next, we consider the effect of being taken POW. For comparability with the other war shocks, we consider a binary indicator, comparing those veterans who were imprisoned for more than six months (a fate shared by nearly half of the men in our data) with those soldiers who were not POWs. Since we drop individuals with short durations of captivity of no more than six months, the number of observations in panel (b) of Table \ref{tab:cs191921} is lower than in previous tables. However, our estimates remain similar if we consider
all POWs regardless of their length of captivity.

POWs born in 1919-21 were imprisoned at a time when--in peacetime--they would have completed their education or entered the labor market. Panel (b) of Table \ref{tab:cs191921} shows that as a result, they received slightly less education (column (1))\footnote{As shown in Online Appendix Table \ref{tab:lifecycle}, the small and insignificant net effect on completed education reflects a significant reduction in educational investments at age 18-25 (during captivity) and increased educational investments at age 26-30 (after returning from captivity).} and were employed for about 2.3 fewer years before age 55 (column (2)). 
They also had less occupational success than veterans who avoided war captivity (column (4)), although this effect is less pronounced when considering the duration of rather than a binary indicator for war captivity (see Online Appendix Table \ref{tab:captivity_continuous}). Despite being employed fewer years, POWs did not receive much lower work pensions (column (5)), as Germany's pension system ``replaces'' employment gaps caused by war captivity (see the appendix for an overview of the pension system).\footnote{However, POWs received significantly lower war victim's pensions (column (6)). This seemingly counterintuitive result reflects that being a POW did not in itself qualify a veteran for a war victim's pension, and that severely wounded soldiers eligible for victim's pension sometimes avoided captivity because they returned home earlier. Accordingly, controlling for war injuries attenuates the effect on war pensions (Table \ref{tab:cs_robustness191921}, see "Extended" specification in panel (b)).} 

Figure \ref{fig:lfeffects_191921}c illustrates how war captivity delayed labor market entrance. POWs were significantly less likely to be employed in their 20s, with the gap peaking at nearly 50 pp in their mid-20s. This gap is unsurprising, reflecting the POWs' forced withdrawal from the labor market due to their imprisonment. But remarkably, POWs managed to close the employment gap in their 30s and were \textit{more} likely to work in later life: POWs' probability of employment overtook that of non-POWs at age 55. The gap widens until age 59, before closing again as increasingly many veterans retire. Conditional on employment, POWs experienced lower occupational success than non-POWs at earlier ages, reflecting their delayed career start, but the gap closes gradually over time (panel d).\footnote{Related, \cite{Maas1999} document that the length of military service, including periods of imprisonment, lowers occupational status in 1950, but that the effect dissipates as the German economy recovers.}

These effects of imprisonment on educational attainment and retirement decisions are again in line with the predictions from standard life-cycle theory, as we show in Online Appendix \ref{sec:theory_captivity}. By reducing the potential duration of an individual's productive working span, imprisonment discourages educational investments. Moreover, by reducing labor earnings, war captivity increases the marginal utility of consumption and, therefore, participation--POWs postpone their retirement entry to compensate for the lost working time during captivity. This is akin to an income effect in static or dynamic models of labor supply \citep[for example][]{Imbens2001}.\footnote{However, the German pension system adjusts for time spent in captivity when calculating pension payments, diminishing this income effect (although it does not adjust for the potential effect of captivity on subsequent earnings).} 

The mechanisms we highlight are therefore distinct from mechanisms related to the potential health effects of imprisonment, which have received much attention in the related literature (see \citealt{myrskyla2023look}). Indeed, the health trajectories of POWs and non-POWs are not very different in our setting (see Online Appendix Figure \ref{fig:injury_illness}b). However, the conditions soldiers experienced during captivity may have additional implications for labor market outcomes. The conditions and duration of captivity varied systematically by the location of imprisonment, and this location is recorded in our data for a subset of the GHS-2 sample. We find that time spent in war captivity had a more negative effect on employment for soldiers who were imprisoned in Eastern Europe or the USSR than for those imprisoned elsewhere (see Online Appendix Table \ref{tab:captivity_continuous}).{\footnote{Since the duration of captivity varies systematically across locations, we here consider a continuous rather than a binary measure of war captivity. To compare the coefficients to our baseline specification, note that the average duration of captivity was 33 months (among those soldiers who spent at least 6 months in captivity).}

\paragraph{Displacement.} More than one in five men in our sample were forcibly displaced from Eastern Europe. As one of the largest population movements in history, the mass arrival of displaced persons in West Germany had important aggregate effects \citep[e.g.][]{BraunKvasnicka2014,BraunWeber2021,Ciccone2022,Peters2022}. However, our interest here is in the (relative) individual economic performance of the displaced, which was of central policy interest in postwar Germany \citep{Bauer2013}. 

We again focus on the 1919-21 birth cohort observed in the GHS. This cohort was in their mid-20s at the time of displacement (we study the role of age-at-displacement below). Panel (c) of Table \ref{tab:cs191921} shows that between the ages of 20 and 55, displaced persons were employed for about 0.6 years less than their non-displaced peers (column (2)), but their employment at older ages was similar (column (3)). Consistent with its negative impact on employment, displacement reduced the maximum occupational prestige attained over a lifetime by 2.4 points (or 5\% relative to the mean of the control group). 

Displaced persons also had significantly lower incomes in old age than non-displaced persons, even though West German pension law equated periods of employment before displacement from Eastern Europe with periods in West Germany \citep{Bauer2019}. This is primarily due to lower nonpension income from rental income, interest, or dividends, which fell by DM 310 per month, or almost 68\%, relative to the control mean (Column (7)). This result is consistent with previous findings that the displaced never fully caught up in terms of wealth \citep{Bauer2013}. 
The decline in non-pension income is especially pronounced among displaced individuals whose fathers had higher levels of education (see Online Appendix Table \ref{tab:displacement_prewarSES}). 

Panels (e) and (f) of Figure \ref{fig:lfeffects_191921} show that displacement left clear traces in labor market careers. Immediately after the war, displaced persons were up to 10 pp less likely to be employed than non-displaced men (panel (e)). The gap in employment decreased in the early 30s and disappeared by the mid-30s. However, conditional on employment, the occupational prestige of displaced persons remained significantly lower throughout their labor market careers. Panel (f) illustrates that this gap was greatest around age 25, shortly after displacement. While this penalty declined in the late 20s, it remained largely unchanged thereafter. At age 56-65, displacement still lowered occupational prestige by around 3.4 points (or 7.8\% relative to the control mean). This aligns with \cite{Bauer2013}, who show that displaced men born in 1906-25 earned 5.1\% less in 1971 than otherwise comparable non-displaced men. Our results suggest that such occupational disadvantages persisted until late in the career, even for individuals who were just starting their careers at the time of displacement.

In Online Appendix \ref{sec:theory_displacement}, we interpret displacement as a decline in the wage rate (for example, due to a loss of social and job networks) or a loss of wealth (for example, due to lost property). A wage decline generates opposing income and substitution effects and, therefore, ambiguous implications for the retirement decision. In a simple model with log-linear utility, the income and substitution effects on labor market entrants cancel out exactly. On the other hand, a pure wealth effect would generate an income but no substitution effect and therefore delay retirement. These implications align with the observation that in the 1919-21 cohort, expellees do not retire (significantly) later. However, as we show below, the effects of displacement depend critically on the age at which a person was displaced, with strong effects on education for younger birth cohorts (whose educational investments were directly interrupted) and large effects on employment for older cohorts (who experience weaker income effects).

\paragraph{Robustness checks and additional evidence.} Table \ref{tab:cs_robustness191921} shows that our baseline results from Table \ref{tab:cs191921} are robust to alternative control variables and to the joint inclusion of all shocks. We present three sets of regressions for each shock. The ``raw'' specification controls only for year of birth. The ``baseline'' specification adds our standard controls: years of schooling of father and mother, number of siblings, and time of war entry. Finally, the ``extended'' specification controls for own years of secondary schooling, an indicator for poor health before age 18, and the respective other shocks. 

Our coefficients of interest remain stable as controls are added and model fit improves, suggesting that selection is unlikely to drive our results. To formalize this argument, we report bias-corrected estimates based on \cite{Oster2019}, which adjust for potential selection on unobservables based on the selection on observables.\footnote{We compute the bias correction in our ``extended'' specification, assuming that selection on observables is proportional to that on unobservables. Following \cite{Oster2019}, we assume that the maximum R2 accounted for by both observable and unobservable confounders is 1.3 times larger than the R2 explained by observables alone.} These estimates are consistent with our baseline estimates, especially with respect to the effects of war shocks on years of employment, occupational prestige, and pension income. The effects on education vary more with the set of controls. In the next section, we take advantage of the large sample size of the 1970 census to present precise estimates of the displacement effects on education by cohort.
 
\begin{table}[!t]
	\centering
	\begin{threeparttable}
		\caption{Robustness tests on the effect of different war-related shocks, men born 1919-21} \centering
		\label{tab:cs_robustness191921}
		\begin{footnotesize}
			\begin{tabular}{lccccccc}
				\toprule
				&Educational& \multicolumn{2}{c}{Years in employment} & Occup.  &  \multicolumn{3}{c}{Old age income from}  \\ \cline{3-4} \cline{6-8}
				&attainment&  age  & age & prestige  & work  & war victim & nonpension\\
				&(years)& 20-55 & 56-65 & (maximum) & pension  & pension & sources \\
				& (1)	& (2)	& (3)  & (4) & (5) & (6) & (7)  \\
				\midrule
				\multicolumn{3}{l}{\textbf{(a) War injury (0/1)}}  & & & & & \\
				Raw & -0.434** &      -0.031   &      -1.008***&      -0.458   &    -263.98** &     195.57***&      23.89   \\
				&     (0.204)   &     (0.307)   &     (0.272)   &     (0.957)   &   (122.75)   &    (51.58)   &   (134.43)   \\
				Baseline &      -0.263   &       0.060   &      -0.908***&      -0.086   &    -233.91*  &     180.63***&     -14.98   \\
				&     (0.199)   &     (0.320)   &     (0.290)   &     (0.987)   &   (122.80)   &    (48.76)   &   (131.74)   \\
				Extended &      -0.080   &      -0.162   &      -0.810***&       0.611   &    -174.14   &     186.78***&      38.12   \\
            			&     (0.153)   &     (0.312)   &     (0.304)   &     (0.932)   &   (126.48)   &    (46.81)   &   (148.85)   \\
				\vspace{-0.3cm} \\
				\textit{Oster}       &       0.011   &      -0.247   &      -0.773   &       0.900   &    -143.20   &     174.41   &      55.68   \\
				\midrule
				\multicolumn{3}{l}{\textbf{(b) War captivity ($>$ 6 months)}}  & & & & & \\
				Raw &      -0.413*  &      -2.286***&       0.386   &      -3.396***&      35.07   &    -131.28***&     -88.68   \\
				&     (0.237)   &     (0.318)   &     (0.324)   &     (1.170)   &   (125.36)   &    (44.91)   &   (133.53)   \\
				Baseline &      -0.262   &      -2.266***&       0.446   &      -2.879** &     -11.71   &    -121.35***&     -62.98   \\
				&     (0.224)   &     (0.332)   &     (0.332)   &     (1.201)   &   (129.94)   &    (42.21)   &   (137.43)   \\				
				Extended &      -0.263   &      -2.102***&       0.542   &      -2.227*  &      18.54   &    -121.01***&     -24.49   \\
			         &     (0.171)   &     (0.352)   &     (0.346)   &     (1.186)   &   (133.22)   &    (39.56)   &   (149.29)   \\	
				\vspace{-0.3cm} \\				
				\textit{Oster}   &      -0.259   &      -2.063   &       0.528   &      -2.159   &      12.29   &    -115.59   &     -28.56   \\
				\midrule
				\multicolumn{3}{l}{\textbf{(c) Displacement (0/1)}}  & & & & & \\
				Raw  &      -0.087   &      -0.854** &      -0.074   &      -2.351** &     -99.85   &      39.58   &    -303.93***\\
				&     (0.233)   &     (0.356)   &     (0.286)   &     (0.990)   &   (131.29)   &    (44.81)   &    (78.99)   \\
				Baseline  &      -0.150   &      -0.638*  &       0.121   &      -2.424** &     -80.08   &      19.74   &    -310.05***\\
				&     (0.233)   &     (0.367)   &     (0.314)   &     (1.015)   &   (144.09)   &    (41.30)   &    (94.47)   \\				
				Extended  &      -0.266   &      -0.472   &       0.104   &      -2.698***&    -105.97   &      11.02   &    -323.62***\\
           			&     (0.163)   &     (0.355)   &     (0.306)   &     (0.927)   &   (136.48)   &    (36.47)   &    (97.30)   \\
				\vspace{-0.3cm} \\							
				\textit{Oster} & -0.301   &   -0.400   &   0.097   & -2.853   &  -116.31   & 7.87   &  -323.23   \\
				\bottomrule 
			\end{tabular}%
		\end{footnotesize}
		\begin{tablenotes}[flushleft]
			\item \footnotesize{\emph{Notes}: Estimates of the effect of war-related shocks on various outcome variables (shown in the table header). Each estimate is from a separate regression. The sample consists of males born 1919-21. Columns (5) to (7) restrict the sample to the second part of GHS-2 conducted in 1987/88 where we observe individuals up to age 65-69 (see Footnote \ref{fn:lvs}). The ``raw'' specification controls only for birth year indicators. The ``baseline'' specification additionally controls for years of schooling of father and mother, number of siblings, and time of entry into the war. The ``extended'' specification additionally controls for own years of secondary schooling, an indicator for poor health before age 18, and all other war shocks. The ``Oster'' specification adjusts for potential selection on unobservables based on \cite{Oster2019}. Robust standard errors in parentheses. ***, **, and * denote statistical significance at the 1\%, 5\%, and 10\% level.}
            \item \footnotesize{\emph{Source}: \cite{ZA2647, ZA2646}}
		\end{tablenotes}
	\end{threeparttable}
\end{table}

Finally, Online Appendix Table \ref{tab:lifecycle} reports additional evidence on the effects of wartime shocks on education, marital status, and number of children over the life cycle. The results are from separate regressions for different age groups (18-25, 26-30, 31-40, 41-55, 56-65), controlling for our standard set of control variables.\footnote{The tables also report age-specific effects on employment and occupational prestige, complementing the detailed life-cycle graphs in Figure \ref{fig:lfeffects_191921} (which only control for year of birth). As can be seen, the life-cycle patterns reported earlier are robust to controlling for parental education, number of siblings, and time of entry into the war.} 
Two findings stand out. First, the different war experiences had large short-term but virtually no long-term effects on marital status or the number of children. War-injured veterans were more likely to be married in their late twenties (up 8.8 pp from a baseline of 49.5\%), presumably because they returned from the battlefield earlier or were more dependent on a partner for help. On the other hand, POWs were significantly less likely to be married in their late 20s (by 23.2 pp). However, for veterans born in 1919-21, neither war injury nor captivity significantly affected marital status  after age 30. The war led to an acute shortage of men and thus strengthened men's bargaining power \citep[see, e.g.,][]{Bethmann2013,Kesternich2020,Battistin2022}, even if they returned from the war disabled or after long captivity. Second, the table also shows that some POWs sought further education after returning from captivity. Thus, POWs in their thirties were more likely than non-POWs to be in education. However, this was not enough to close the negative educational gap created by their captivity in their early 20s.

\section{The Effect of Displacement across Cohorts}\label{sec:section5}

The previous section examined how individual war experiences shaped labor market careers within a given birth cohort. We now examine how the impact of individual war experiences varies across cohorts. We focus on the impact of displacement, which affected young and old at very different points in the life cycle, on educational attainment and labor market exit. Our analysis relies mainly on the 1970 census, whose large sample size allows us to identify displacement effects by cohort with precision.

\paragraph{Education.} Figure \ref{fig:displ_education_main} compares, separately for men and women, the average educational attainment of displaced and non-displaced persons for cohorts born 1905-1943. We exclude more recent cohorts as they may not have completed their education by 1970. 

Among men, we observe nearly identical trends in education for displaced and non-displaced persons born before 1920. Beginning with cohorts born in the early 1920s, the average educational attainment of the displaced declined, while it stagnated for the non-displaced. Displacement, therefore, slightly reduced the education of men who were in their early 20s at the time of expulsion. This educational penalty gradually increased in subsequent cohorts, reaching a maximum of about 0.7 years for men born around 1930. These cohorts suffered from the turmoil of flight and expulsion during the transition from school to vocational training.\footnote{Unreported regressions confirm that there was a sharp decline in the duration of apprenticeship among displaced persons born around 1930. The average duration of apprenticeship was 0.43 years less for displaced persons born in 1930 than for their non-displaced peers. This gap explains almost 2/3 of the total education gap of -0.68 years.}

The gap between displaced and non-displaced males shrinks for even younger cohorts and in fact turns positive for cohorts who entered school after displacement, in line with earlier results for the cohort born 1944-49 \citep{Bauer2013}. Such positive effects could result from the loss of land, which led the children of displaced persons to seek work outside of agriculture, thereby increasing the importance of educational investment \citep{Bauer2013}. Moreover, the experience of losing property might have encouraged displaced people to invest in ``portable assets'' such as education \citep{Brenner1981,Becker2020}.

\begin{figure}[!t]
	\caption{The impact of displacement on education}\label{fig:displ_education_main}
	\centering
	\begin{threeparttable}
		\begin{center}
		\includegraphics[width=0.75\textwidth]{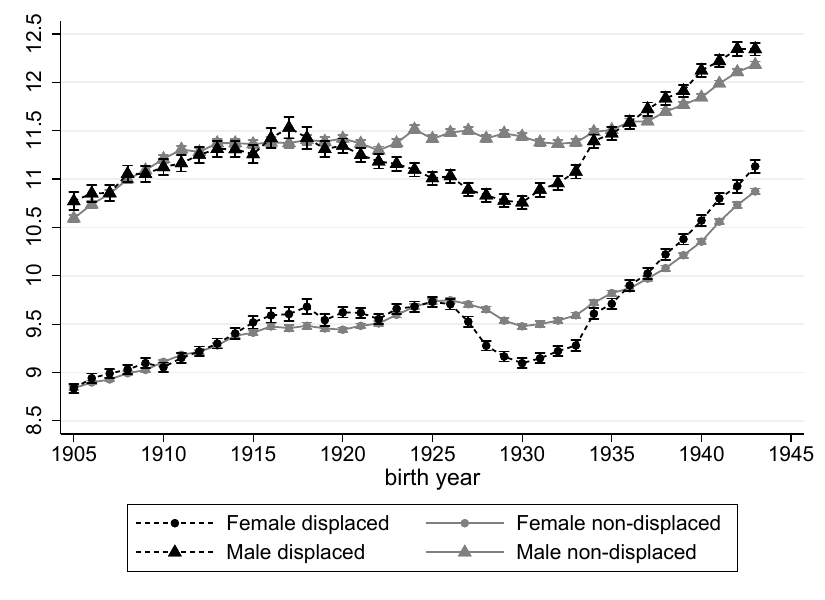}
		\end{center}
		\begin{tablenotes}[flushleft]
			\item \footnotesize{\emph{Notes}: The figure illustrates the impact of displacement on education across cohorts. It shows unconditional means in years of education by cohort and displacement status.}
            \item \footnotesize{\emph{Source}: \cite{RFSOSO}}
		\end{tablenotes}
	\end{threeparttable}
\end{figure}

Similar patterns emerge for women: Displaced women born around 1930 suffered educational losses of about 0.4 years, while women displaced at young age had higher educational attainment than their non-displaced counterparts. The gap between the displaced and the non-displaced opened later for women than men, around the 1926/27 cohort, perhaps because most women in earlier cohorts had completed their education when the war began (as they had, on average, lower education than men). Moreover, women's educational careers were less directly interrupted by the war, while military service forced men to postpone further educational investments. 

One drawback of the 1970 census data is that we cannot control for parental background. However, we observe similar patterns in the GHS in regressions that condition on parental education (see Online Appendix Table \ref{tab:displacement_educ_LVS}): the effect of displacement on educational attainment was moderately negative (positive) for males (females) born in 1919-21, strongly negative for both sexes in the 1929-31 cohort, and positive (but non-significant) for those born in 1939-41, who started their school careers after displacement. We also verify in the GHS that the educational gap between displaced and non-displaced in the 1929-31 cohort opens only with displacement (see Online Appendix Figure \ref{fig:displ_education}).

Beyond the displaced, Figure \ref{fig:displ_education_main} also illustrates that the secular expansion of education was halted and reversed for cohorts born around 1930. These cohorts suffered from school closures and a lack of apprenticeships in the postwar years \citep{Mueller2004}. Our results thus also testify to the cohort-wide educational costs of WWII, documented in \cite{Ichino2004}.

\paragraph{Employment.} Displacement had little impact on the retirement behavior of men who experienced WWII as young adults (see previous section), but how did it affect women or older cohorts? We use an individual's last employment to define an indicator that takes the value of one for years in which an individual has left gainful employment (``employment exit''). 

Figure \ref{fig:displ_empl_education_exit_cohort} illustrates how the probability of exiting employment evolves over the life cycle, distinguishing between displaced and non-displaced persons. We consider four cohorts born in 1885, 1890, 1895, and 1900. For all four cohorts, and both genders, employment exits of displaced persons spiked in 1945 when most displacements occurred (the second gray vertical line in the figures indicate the corresponding ages of 60, 55, 50, and 45). In contrast, there is no spike for the non-displaced, for whom the exit probability evolves smoothly around 1945.\footnote{For men, the plot also shows a second jump at the statutory retirement age of 65.} The employment gap between the groups, therefore, widened sharply at displacement. Many expellees never returned to the labor market after losing their jobs during the displacement.

\begin{figure}[!t]
	\caption{Employment exit probability over the life cycle, by cohort}\label{fig:displ_empl_education_exit_cohort}
	\centering
	\begin{threeparttable}
		\hspace*{-0.6cm}\subfloat[1885]{\includegraphics[width=0.525\textwidth]{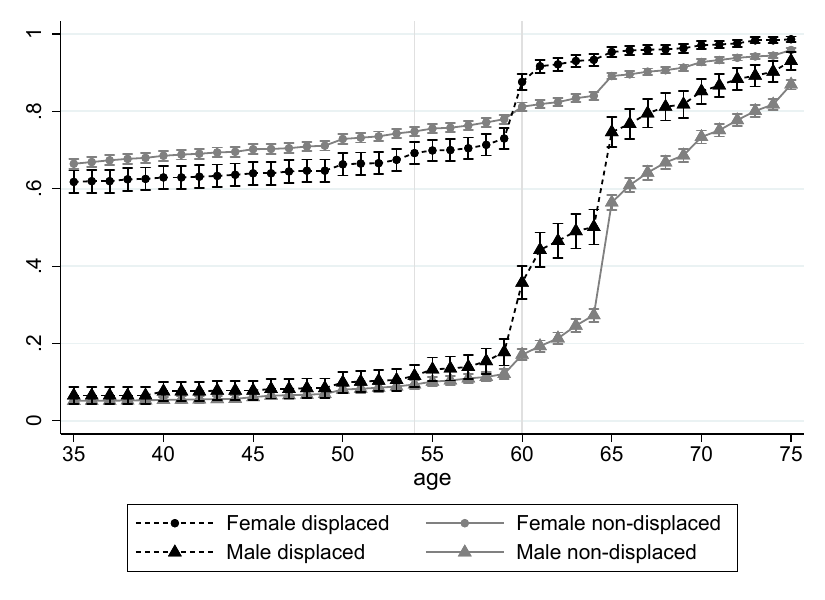}} 
		\subfloat[1890]{\includegraphics[width=0.525\textwidth]{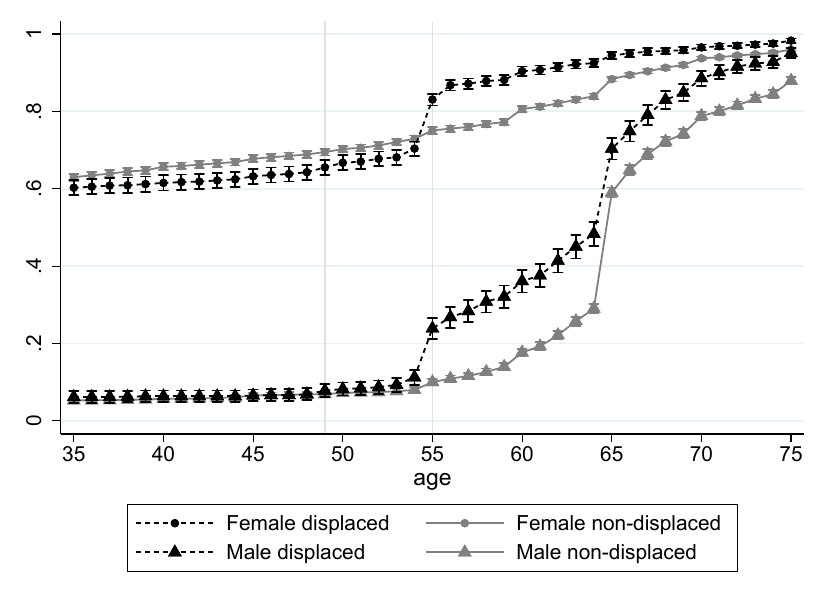}} \\
		\hspace*{-0.55cm}\subfloat[1895]{\includegraphics[width=0.525\textwidth]{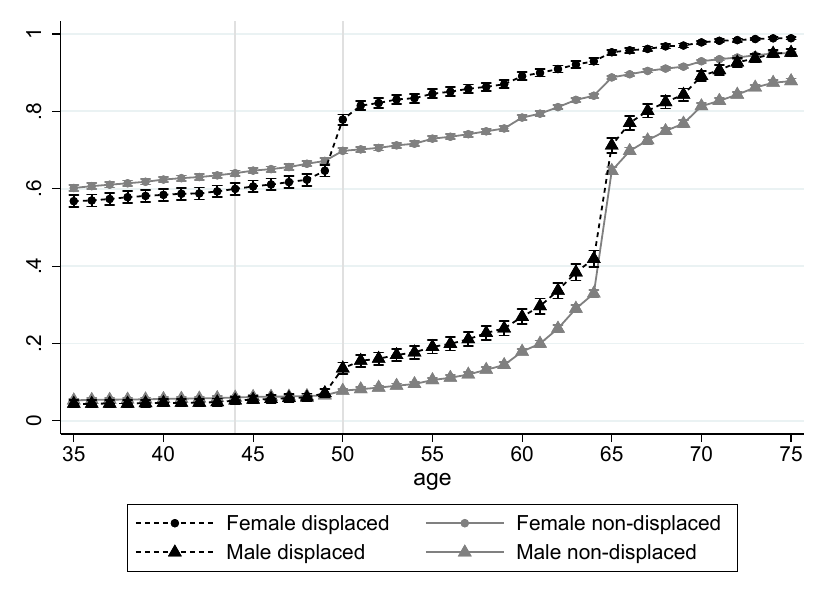}} 
		\subfloat[1900]{\includegraphics[width=0.525\textwidth]{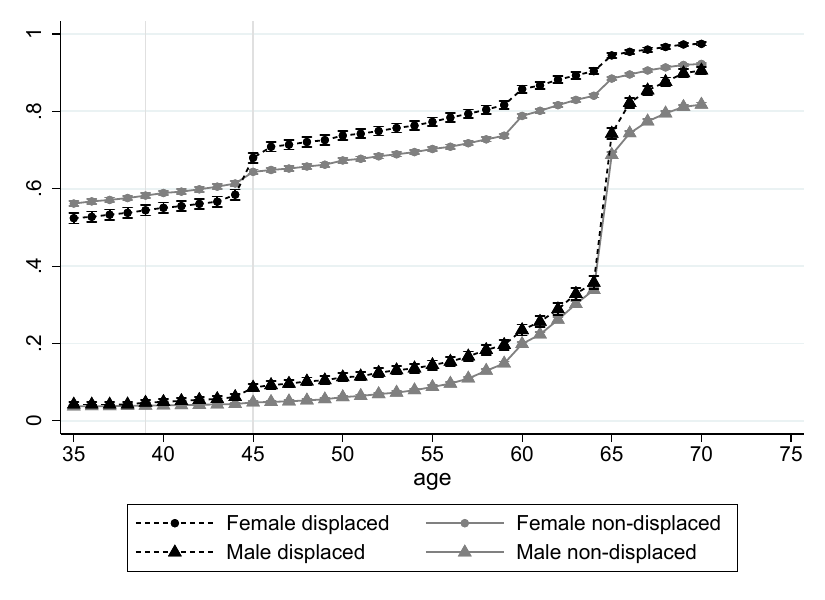}}
		\begin{tablenotes}[flushleft]
			\item \footnotesize{\emph{Notes}: The figures depict, by cohort, the probability of having exited employment for displaced and non-displaced persons over the life cycle. Gray vertical lines indicate the beginning and end of WWII. Vertical bands indicate 95\% confidence interval.}
            \item \footnotesize{\emph{Source}: \cite{RFSOSO}}
		\end{tablenotes}
	\end{threeparttable}
\end{figure}

Importantly, this effect of displacement on employment varies greatly by birth cohort and gender. Panel (a) of Figure \ref{fig:displ_empl_exit} illustrates this pattern in detail, reporting the immediate employment effect of displacement for each cohort born between 1880 and 1905. Estimates are from simple Difference-in-Differences (DiD) regressions, which measure the difference between displaced and non-displaced individuals in the change in the labor market exit probability between 1938 (pre-) and 1946 (post-treatment period). Specifically, we estimate, separately for each cohort, the following regression equation 
\begin{equation}	
	exit_{it} = \alpha + \beta displaced_i + \gamma d_t + \delta (displaced_i \times d_t) + \varepsilon_{it},
\end{equation}
where $exit_{it}$ indicates whether individual $i$ has left the labor market by year $t$, $displaced_i$ is a dummy variable for (to be) displaced persons, and $d_t$ is an indicator for the post-displacement year 1946. The parameter of interest is $\delta$.

\begin{figure}[!t]
	\caption{The impact of displacement on employment}\label{fig:displ_empl_exit}
	\centering
	\begin{threeparttable}
		\hspace*{-0.6cm}\subfloat[Immediate effect on employment exit]{\includegraphics[width=0.525\textwidth]{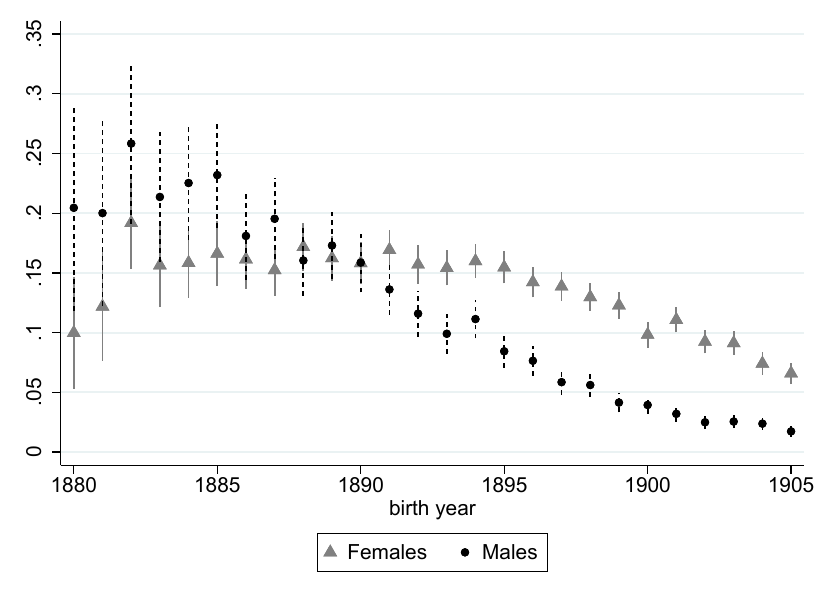}} 
		\subfloat[Total employment effect (years)]{\includegraphics[width=0.525\textwidth]{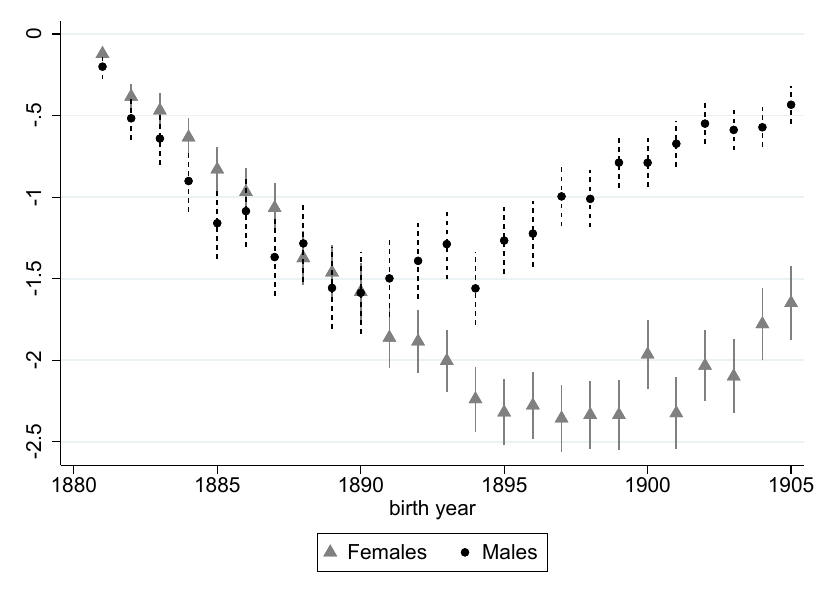}}
		\begin{tablenotes}[flushleft]
			\item \footnotesize{\emph{Notes}: The figure illustrates the impact of displacement on employment across cohorts. Panel (a) estimates the immediate effects of displacement on the probability to exit from employment for cohorts born between 1880 and 1905. Effect estimates are from DiD regressions, with 1938 as the pre- and 1946 as the post-treatment period. Panel (b) estimates the overall impact of displacement on years of employment up to age 65. The effect estimates are from DiD regressions, again with 1938 as the pre-treatment period. The post-treatment period extends from 1946 to the year a cohort turns 65. Point estimates are indicated by a dot, vertical bands indicate 95\% confidence intervals.}
            \item \footnotesize{\emph{Source}: \citep{ RFSOSO}}
		\end{tablenotes}
	\end{threeparttable}
\end{figure}

For men born in 1885, displacement increased the probability of employment exit in 1946 by 23.2 pp (relative to a post-treatment control mean of 19.3\%). The effect gradually declines for more recent cohorts who were displaced at younger age, to less than 1.7 pp for men born in 1905. The immediate effects of displacement on employment are more stable with age for women. For older women born in 1885, displacement increased the probability of exiting the labor market by 16.6 pp in 1946. The effect size is substantial, given that only a third had not yet left the labor market by 1938. Thus, half of the women still ``at risk'' of exiting did so due to displacement. Displacement also had a much greater effect on younger women than on men: Among women born in 1905, 6.9 pp left employment permanently by 1946 as a result of displacement. Presumably, the displacement effect is larger for women than men because the former were on average less attached to the labor market at the time.

While the immediate impact is smaller for younger cohorts, the young have a longer labor market career ahead of them. An earlier employment exit is, therefore, more consequential for them. Figure \ref{fig:displ_empl_exit}b depicts the total effect of displacement on years of employment up to age 65, the statutory retirement age. We quantify the total employment effect as the cumulative difference in employment from 1946 to the year a cohort turns 65.\footnote{Specifically, we first estimate the average effect of displacement on employment in the post-treatment period in a DiD design, with displaced workers as the treatment group, 1938 as the pre-treatment period, and the post-treatment period extending from 1946 to the year in which a cohort turns 65. We then multiply this average effect by the length of the post-treatment period, that is, the potential years of employment before a worker turns 65).} For men, the total effect of displacement follows a hump shape: Older cohorts lost little time in employment because they were close to retirement anyway. The employment loss then gradually increases with birth year, peaking at about 1.5 years for men born in 1890-95. Younger male cohorts again lost very little employment time, as only a few exited employment after displacement. For women, on the other hand, the overall employment loss is largest for the relatively young cohorts born in 1895-1905. They lost more than two years of employment due to displacement.

In Online Appendix \ref{sec:theory}, we show that the observed variation in the employment effect is consistent with simple theoretical arguments. A reduction in the wage rate due to displacement generates a negative substitution effect (as work is being less rewarded) and a positive income effect (as life-cycle earnings and consumption decrease). However, this income effect depends on the age at which an individual is being displaced. Individuals close to their expected retirement age experience only a minor income effect, as most of their life-cycle earnings have already been realized--their employment response is dominated by the substitution effect, and hence negative. In contrast, younger individuals experience a more sizable income effect due to displacement, muting the response in employment.

\section{Conclusion}\label{sec:conclusion}

The dramatic return of war and displacement to Europe following Russia's attack on Ukraine has reignited interest in the labor market consequences of violent conflict. Our study examines the impact of battlefield injuries, war captivity, and displacement in the context of WWII, the most devastating conflict in history. We show that the economic consequences of these shocks often became visible long after the war. For example, the effect of war injuries on the employment of WWII veterans was most pronounced not in the immediate postwar period, but decades later, as these veterans approached retirement. Displacement also had very different effects depending on the age and gender of the displaced. The impact on education was worst for adolescents facing the critical transition from school to vocational training. On the other hand, the loss in employment was particularly severe for women and older male cohorts. Overall, our findings suggest that policies to alleviate the hardships of war should take into account that its consequences depend critically on age-at-exposure and vary greatly over the life cycle. 

Our results align with standard life-cycle theory, suggesting relevance beyond WWII. However, the precise magnitudes of the effects are likely to vary with labor market conditions and institutional support. In postwar West Germany, strong economic growth and low unemployment may have facilitated the integration of war-disabled veterans, POWs, and displaced persons, and a supportive pension system may have influenced their retirement decisions. An open question for future research is whether life-cycle patterns similar to those in postwar Germany hold in contexts with less favorable postwar conditions and weaker institutional support. 

\section*{Appendix: Pension Benefits and World War II}

In what follows, we describe the provisions of the pension system as they were relevant to the 1919-21 birth cohort (see Online Appendix \ref{subsec:pensions} for additional details). Statutory pensions in Germany depend on the labor income earned over the life course. The longer people work and the more they earn, the higher their pensions. Since the pension reform of 1957, the system has been organized as a pay-as-you-go-scheme, in which current contributions pay for current pension obligations.

 \paragraph{Old-age pensions.} Before 1992, the statutory retirement age for old-age pensions was 65. Thus, regular retirement was open to anyone at age 65 who had contributed to the pension system for at least five years. Those with at least 35 years of contributions could retire at age 63 under the flexible retirement option introduced in 1972. Retirement at age 60 was possible under certain conditions for women, the long-term unemployed, and the disabled (see Online Appendix \ref{subsec:pensions} for details).

The introduction of flexible retirement in 1972 led to a significant decline in the average entry age for old-age pensions \citep{BoerschSupan1998}, especially for men. Their average age of first claiming an old-age pension fell from 65.1 years in 1972 to a low of 62.3 years in 1982 and thus well below the standard retirement age of 65. The distribution of retirement ages peaked at ages 60, 63 and 65 \citep{BoerschSupan1998}. These peaks correspond to the earliest ages at which early retirement for health and labor market reasons, flexible retirement for workers with long work histories, and regular retirement for workers with short work histories were possible. 

 \paragraph{Disability and survivor pensions.} In addition to old-age pensions, the pension system provides disability benefits for workers of any age, subject to medical indication and a minimum insurance period of at least five years. Disability benefits came in two forms. General disability benefits  (\textit{Erwerbsunf\"{a}higkeitrente}) were for those unable to work regularly and made up 80\% of approved claims in the late 1980s \citep{Riphahn1999}. Occupational disability benefits  (\textit{Berufsunf\"{a}higkeitrente}) were for those with less than half the work capacity of a healthy worker in the same job and were one-third lower than full old-age pensions.

Disability benefits have been an important route to retirement in Germany. Throughout the late-1970s and 1980s, the share of workers retiring on disability benefits was higher than the share retiring on regular old-age pensions \citep{BoerschSupan1998,Riphahn1999}. Those retiring on disability pensions in the 1980s were on average in their mid-fifties.  Consequently, the average age of all new retirees (including those retiring on old-age pensions) in Germany fell below 59 in the early 1980s.

The pension system also covers the financial loss caused by the death of a spouse. The survivor's pension  (\textit{Witwenrente}) is intended to replace the support previously provided by the deceased. 

 \paragraph{Gaps in employment biographies.} Importantly, the pension system smooths out gaps in the employment biography caused by compulsory state measures such as military service, war captivity, expulsion, and resettlement. These "substitute periods" (\textit{Ersatzzeiten}) are fully considered when calculating the pension. In addition, "periods of absence" (\textit{Ausfallzeiten}) are considered in the pension calculation. Periods of absence are periods during which employment is interrupted for personal reasons, including unemployment, incapacity to work, pregnancy, and further education.


\singlespacing
\bibliographystyle{apalike}
\begin{small}

\end{small}
\newpage

\appendix
\renewcommand{\thetable}{\Alph{table}}
\renewcommand{\thetable}{\Alph{section}\arabic{table}}
\renewcommand{\thefigure}{\Alph{section}\arabic{figure}}
\renewcommand{\theequation}{\Alph{section}-\arabic{equation}}
\setcounter{page}{1}
\setcounter{equation}{0}
\setcounter{table}{0}
\setcounter{figure}{0}

\section*{Online Appendix}

\section{Data Appendix}

\subsection{\label{sec:Education-Measure}Educational attainment}

The GHS indicates the highest school-leaving and vocational training qualifications that a person has obtained (if any). Using this information, we calculate years of schooling as the minimum duration required to earn a particular degree. To determine the total number of years of education, we add to the years of schooling the minimum length of time required to earn a particular vocational education degree. Table \ref{tab:Lengths-Degree} shows the minimum length of time we use to calculate our measures of education \citep[taken primarily from][]{Muller1979}. 

\begin{table}[h]
	\begin{centering}
		\caption{\label{tab:Lengths-Degree}Minimum lengths of time required
				to earn a given degree}
		\par\end{centering}
	~
	\centering{}%
	\begin{small}
	\begin{tabular}{lc}
		\toprule	
		Degree & Minimum time length \\
		\hline 
		\emph{School Degree} & \\
		No completed school degree & 8 years\\
		Sonderschulabschluss (special needs school) & 8 years\\
		Volks-/Hauptschulabschluss (low school track) & 8 years\\
		Mittlere Reife (medium school track)  & 10 years\\
		Fachhochschulreife (high school track) & 12 years\\
		Abitur (high school track) & 13 years\\
		\hline 
		\emph{Vocational Training Degree} & \\
		No vocational degree & 0 years\\
		Agricultural or household apprenticeship & 2 years\\
		Industrial apprenticeship & 2 years\\
		Vocational school degree & 2 years\\
		Commercial apprenticeship & 3 years\\
		Master craftsman & 4 years\\
		University of applied sciences degree & 4 years\\
		University degree & 5 years\\
		Other vocational training degree  & 2 years\\
		\bottomrule
	\end{tabular}
	\end{small}
\end{table}

\section{Pension Benefits and World War II}\label{subsec:pensions}

In what follows, we describe the provisions of the pension system as they were relevant to the 1919-21 birth cohort.\footnote{See \cite{Allmendinger1994} for further details, especially on the gendered impact of the pension system on this generation's life courses. \cite{Mierzejewski2016} provides a comprehensive history of the German pension system. \cite{Riphahn1999} studies the predictors of the transition into disability retirement of German men in the 1980s.} Statutory pensions in Germany depend on the labor income earned over the life course. The longer people work and the more they earn, the higher their pensions. The German pension system thus ``insures'' living standards achieved during working life and extends prosperity into retirement.  Since the pension reform of 1957, the system has been organized as a pay-as-you-go-scheme. This means that current contributions (from employees and employers) pay for current pension obligations. The 1957 reform also made pensions dynamic by linking them to wage trends. Entitlement to a pension arises when individuals have paid contributions for at least five years. 

\paragraph{Old-age pensions.} Before 1992, the statutory retirement age for old-age pensions was 65. Thus, \textit{regular retirement} was open to anyone at age 65 who had contributed to the pension system for at least five years. Those with at least 35 years of contributions could retire at age 63 under the \textit{flexible retirement} option introduced in 1972. 

Retirement at age 60 was possible under certain conditions for women, the long-term unemployed, and the disabled. Women could retire at age 60 if they had been insured in the public pension system for at least 15 years and had paid contributions for more than ten years after their 40th birthday.\footnote{The 1992 pension reform abolished this gendered path to retirement, which effectively set the statutory retirement age for most women at 60.} The unemployed could retire at 60 if they had paid contributions for eight of the previous ten years and had been unemployed for 12 of the previous 18 months. The severely disabled could also retire at 60 if they had at least 35 years of contributions and an officially recognized disability.

The introduction of flexible retirement in 1972 led to a significant decline in the average entry age for old-age pensions \citep{BoerschSupan1998}, especially for men. Their average age of first claiming an old-age pension fell from 65.1 years in 1972 to a low of 62.3 years in 1982 and thus well below the standard retirement age of 65. The distribution of retirement ages peaked at ages 60, 63 and 65 \citep{BoerschSupan1998}. These peaks correspond to the earliest ages at which early retirement for health and labor market reasons, flexible retirement for workers with long work histories, and regular retirement for workers with short work histories were possible.  

\paragraph{Disability and survivor pensions.} In addition to old-age pensions, the pension system provides \textit{disability benefits} for workers of any age, subject to medical indication and a minimum insurance period of at least five years. Disability benefits took two forms. \textit{General disability benefits} (\textit{Erwerbsunf\"{a}higkeitrente}) were paid to individuals unable to perform continuous, regular employment. This form accounted for 80\% of all approved disability claims in the late 1980s \citep{Riphahn1999}. Disability benefits (\textit{Berufsunf\"{a}higkeitrente}) were paid to individuals whose ability to work was less than half that of a healthy worker in the same job. These benefits were one-third lower than the full old-age pension, based on the assumption that recipients could still engage in marginal gainful activity. 

Disability benefits have been an important route to retirement in Germany.\footnote{Another important route was through pre-retirement schemes. They allowed workers to leave their jobs early and receive severance pay and unemployment benefits before officially retiring at age 60 under the early retirement provisions for unemployed workers.} Throughout the late-1970s and 1980s, the share of workers retiring on disability benefits was higher than the share retiring on regular old-age pensions \citep{BoerschSupan1998,Riphahn1999}. Those retiring on disability pensions in the 1980s were on average in their mid-fifties.  As a consequence, the average age of all new retirees (including those retiring on old-age pensions) in Germany fell below 59 in the early 1980s. 

The pension system also covers the financial loss caused by the death of a spouse. The survivor's pension (\textit{Witwenrente}) is intended to replace the support previously provided by the deceased. Until 1986, women received survivor's pension unconditionally and regardless of their own work history. Widowers, on the other hand, were entitled to a survivor's pension only if the deceased's wife had provided most of the family's support. This differential treatment did not end until 1986.    

\paragraph{Gaps in employment biographies.} Importantly, the pension system smooths out gaps in the employment biography caused by compulsory state measures such as military service, war captivity, expulsion, and resettlement. These ``substitute periods'' (\textit{Ersatzzeiten}) are fully taken into account when calculating the pension. In addition, ``periods of absence'' (\textit{Ausfallzeiten})\footnote{The pension reform of 1992 changed the term to \textit{Anrechnungszeiten}.} are taken into account in the pension calculation. Periods of absence are periods during which employment is interrupted for personal reasons, including unemployment, incapacity to work, pregnancy, and further education. 

\section{Level of and Expenditure on War Victims' Compensation}\label{sec:compensation}  

The war victims' pension (\textit{Kriegsopferrente}) is paid to persons who have suffered serious health damage as a result of military or military-like service in connection with the war (e.g. damages due to direct warfare, captivity, or internment abroad). Figure \ref{fig:levelwarvictim} depicts compensation payments to war-disabled persons in percent of average gross earned income from 1951 to 1980. It distinguishes between basic and equalization pensions and between war-disabled with a reduction in their civilian earning capacity of 30\%, 50\%, 70\% and 100\%. 

\begin{figure}[!t]
	\caption{War victims' pension by reduction in earning capacity (in \% of gross labor income), 1951-80}
	\label{fig:levelwarvictim}
	\begin{threeparttable}
		\begin{center}
			\includegraphics[width=1.0\textwidth, angle=360]{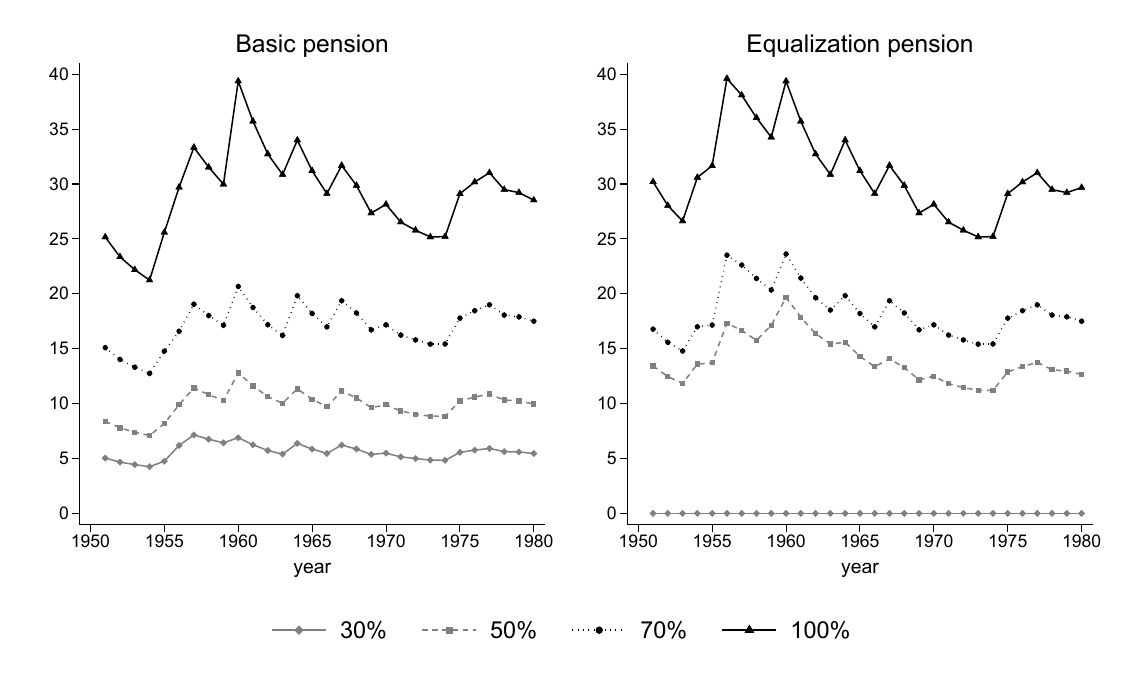}
		\end{center}
		\begin{tablenotes}[flushleft]
			\item \footnotesize{\textit{Notes}: The figure shows compensation payments to war-disabled persons in \% of average gross earned income. It distinguishes between war-disabled persons with a reduction in their civilian earning capacity of 30\%, 50\%, 70\% and 100\%. The left panel shows the basic pension levels, the right panel the maximum compensatory pension levels. See the description in Section \ref{sec:data} and Appendix \ref{sec:compensation} for further details.}
			\item \footnotesize{\emph{Source}: Author's calculations based on the \textit{Bundesversorgungsgesetz} (in its various versions). Data on average gross labor income are taken from \cite{BMJV2020}.}
		\end{tablenotes}
	\end{threeparttable}
\end{figure}

The left panel of Figure \ref{fig:levelwarvictim} shows the level of the basic pension (\textit{Grundrente}). This part of the war victim's pension was not means-tested and was paid to all war victims with a reduction in civilian earning capacity of at least 25\%. Depending on the degree of disability, its level ranged from about 5-7\% to about 25-40\% of gross labor income. The right panel shows the maximum attainable level of the equalization pension (\textit{Ausgleichsrente}). Only war-damaged persons with a reduction in earning capacity of at least 50\%, slightly less than half of all war-damaged persons, were eligible for this part of the war victims' pension. As shown, the maximum amount of the equalization pension was at least as high as the basic pension. However, because the equalization pension was means-tested, few of those eligible received the full amount.\footnote{In 1956, for example, 46.4\% of all war-damaged persons were in principle eligible for an equalization pension because their earning capacity had been reduced by 50\% or more \citep{PIB1957}. However, of those eligible, 63\% received no equalization pension and only 12.6\% received the full amount.}

Figure \ref{fig:levelwarvictim} shows that after a peak in 1960, the level of war victims' pensions relative to average gross earnings declined again. However, additional benefits were introduced in 1960. In particular, severely disabled persons received an additional compensation (\textit{Schadensausgleich}) if their income was below what they would have earned without the war injury. The compensation initially amounted to 30\% and, from 1964, to 40\% of the loss of income. Disabled persons, defined as those whose earning capacity was reduced by 90\% or more, received additional supplements.

\begin{figure}[!t]
	\caption{Social expenditures on war victims (\% of total social expenditures), 1950-80}
	\label{fig:expwarvictim}
	\begin{threeparttable}
		\begin{center}
			\includegraphics[width=0.75\textwidth, angle=360]{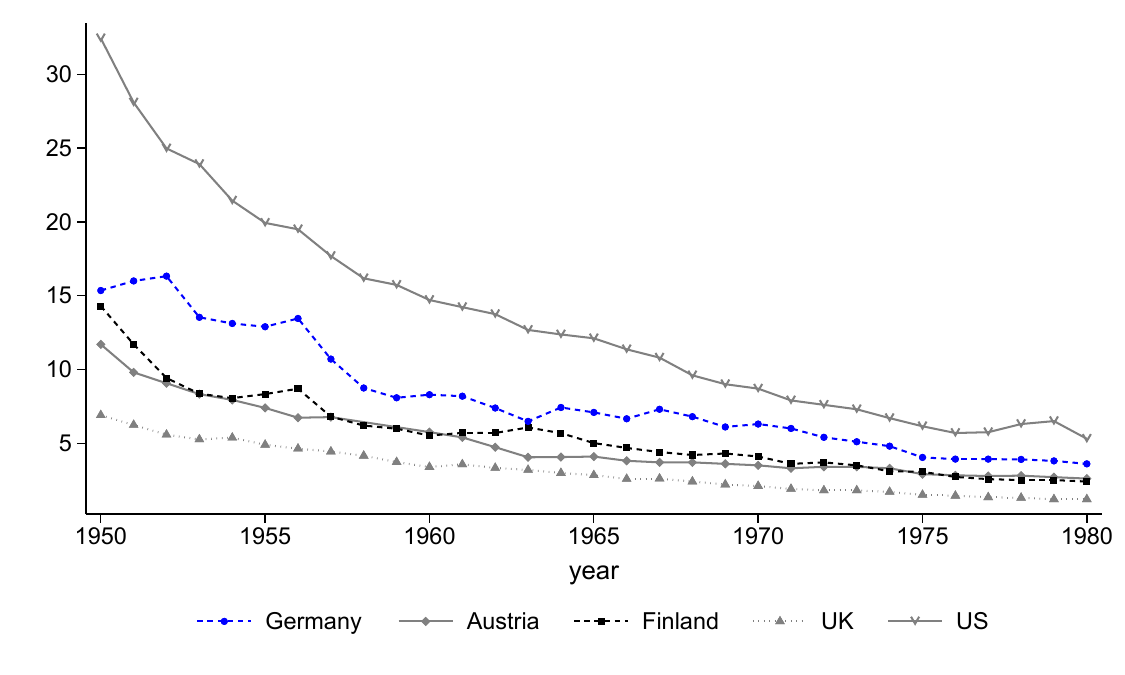}
		\end{center}
		\vspace{0.5cm}
		\begin{tablenotes}[flushleft]
			\item \footnotesize{\emph{Source}: International Labour Organization. The cost of social security. Various volumes.}
		\end{tablenotes}
	\end{threeparttable}
\end{figure}

Total spending on war victims was substantial, especially in the immediate postwar period. \ref{fig:expwarvictim} shows total social spending on war victims as a percent of total social spending from 1950 to 1980. Expenditure shares in Germany are compared with those in Austria, Finland, the United Kingdom and the United States. The figure shows that the share of social expenditure on war victims in Germany hovered around 15\% in the early to mid-1950s and then gradually declined. The decline was initially driven mainly by the decline in the number of orphans eligible for benefits (as the orphan's pension was only paid to minors) and by the deaths of the older cohorts who served in World War I \citep{PIB1957}. 

Germany's share of spending on war victims was comparatively high, reflecting the large number of military deaths. Only in the US was the share even higher, reaching one-third of total social spending. This is largely due to the fact that the welfare state was less developed in the US at the time \citep{Obinger2018}. In addition, unlike in Germany, all veterans were eligible for benefits in the US, not just the war-disabled. Nevertheless, as a percentage of GDP, Germany's spending on war victims in the early 1950s, at 2\%, was about twice that of the US.  

\section{Cohort Effects}
\label{sec:cohorteffects}

Figure \ref{fig:lc_profiles_191951} compares the life-cycle profile of the 1919-21 cohort with later-born cohorts covered by the GHS-1, illustrating that their transition from education to work was dramatically different.  As discussed in the main text, men born in 1919-21 spent, on average, just 156 months in the labor market by age 37, more than 60 months less than males born in 1929-31 or 1939-41. The latter cohorts were either not (1929-31) or only partially (1939-41) conscripted after West Germany rearmed in 1955.  For women we see the inverse pattern: Compulsory work services, unusually high labor demand during war time, and the absence of men accelerated labor market entry of the 1919-21 cohort and led to comparably high participation rates. In their first 37 years of life, females born 1919-21 participated in the labor market twelve and 18 months longer than those born 1929-31 and 1939-41, respectively.

\begin{landscape}
	\begin{figure}[!tbp]
		\caption{Education, participation, and non-participation over the life cycle, by gender and cohort}\label{fig:lc_profiles_191951}
		\centering
		\begin{threeparttable}
			\begin{center}
				\subfloat[Men, education]{\includegraphics[width=0.50\textwidth]{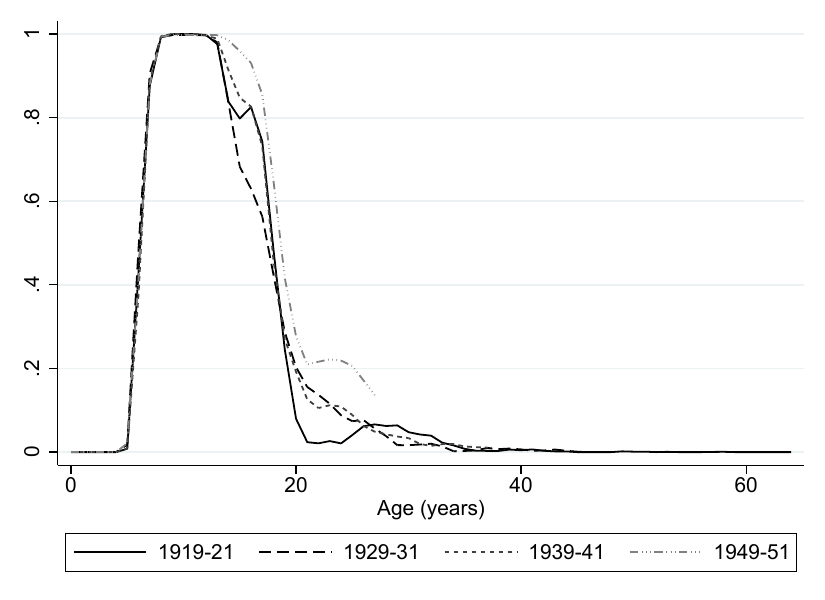}}\hfill
				\subfloat[Men, participation]{\includegraphics[width=0.50\textwidth]{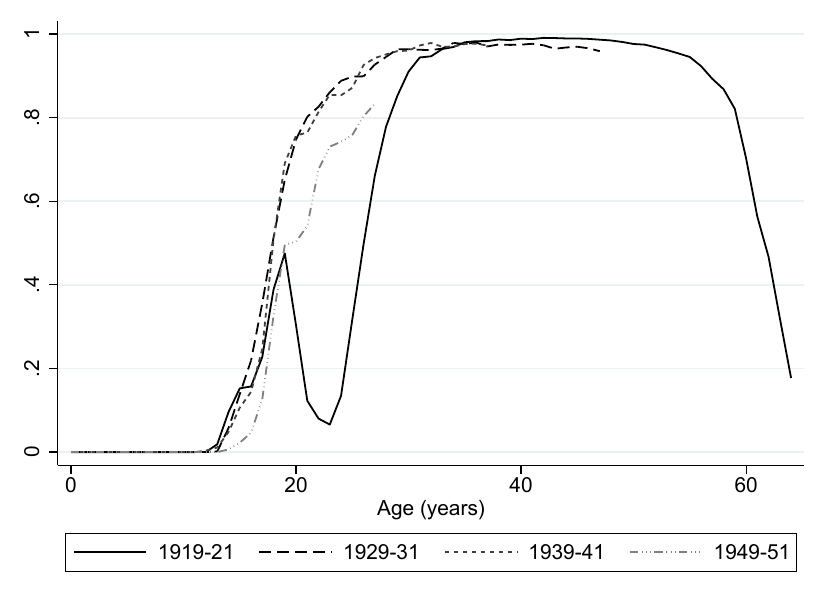}} \hfill
				\subfloat[Men, non-participation]{\includegraphics[width=0.50\textwidth]{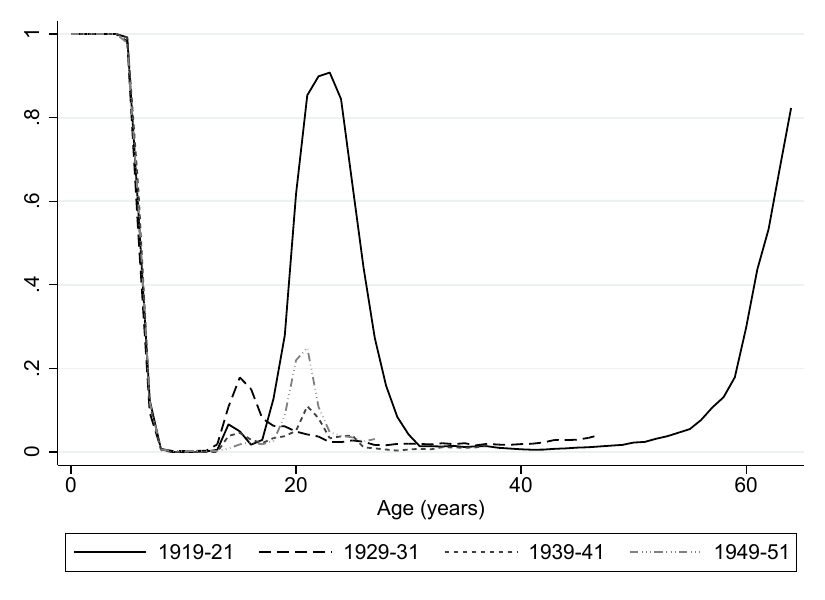}}  \\
				\subfloat[Women, education]{\includegraphics[width=0.50\textwidth]{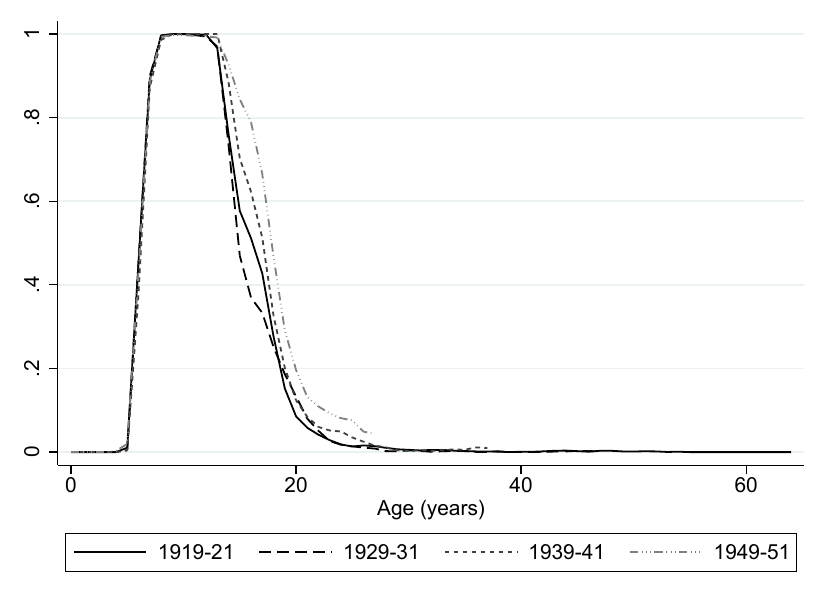}}\hfill
				\subfloat[Women, participation]{\includegraphics[width=0.50\textwidth]{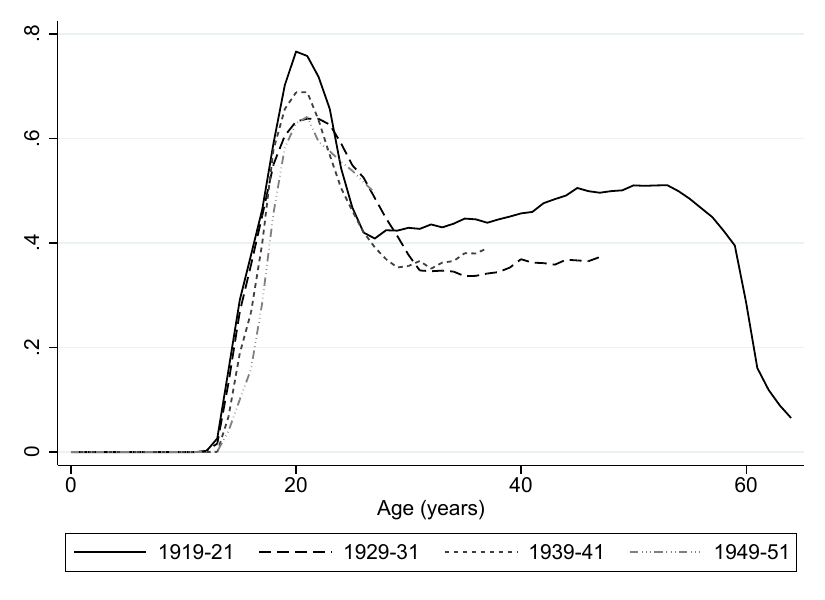}}\hfill
				\subfloat[Women, non-participation]{\includegraphics[width=0.50\textwidth]{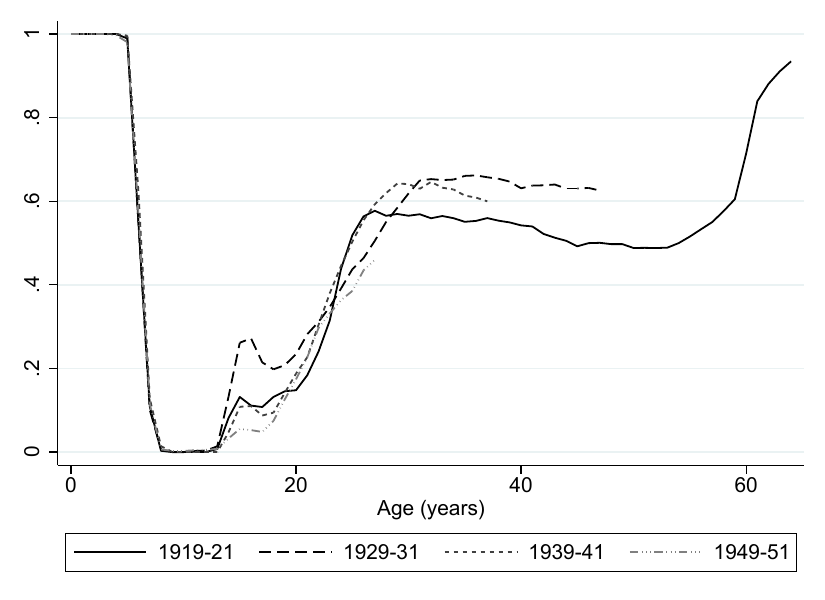}}
			\end{center}
			\begin{tablenotes}[flushleft]
				\item \footnotesize{\emph{Notes}: The graph depicts, separately by gender, the share of individuals in education (Panels (a) and (d)), in the labor force (Panels (b) and (e)) and non-participating (Panels (c) and (e)). We distinguish between cohorts born in 1929-21, 1929-31, 1939-41 and 1949-51. Education includes schooling, vocational training, and further education. Individuals are in non-participation if they are not in education, do not work, and are not unemployed.}
			\end{tablenotes}
		\end{threeparttable}
	\end{figure}
\end{landscape}

\section{Multiple Shocks and Heterogeneous Effects}
\label{sec:multishocks}

\begin{table}[!htb]
	\begin{center}
	\begin{threeparttable}
	\caption{Distribution of war-related shocks} \centering
	\label{tab:corrshocks}
	\begin{footnotesize}		
	\begin{tabular}{cccc}
	\toprule 
	\multicolumn{4}{l}{\textbf{(a) Frequencies}} \tabularnewline
 	War injuries & War captivity & Displacement & Observations \tabularnewline	
	No  & No  & No  & 50  \tabularnewline
	Yes & No  & No  & 33  \tabularnewline
	Yes & Yes & No  & 77  \tabularnewline
	Yes & No  & Yes & 12   \tabularnewline
	 No  & Yes & No & 213 \tabularnewline
	No  & Yes & Yes & 61  \tabularnewline
	No  & No  & Yes & 13  \tabularnewline
	Yes & Yes & Yes &26   \tabularnewline	
	\midrule
	\multicolumn{4}{l}{\textbf{(b) Correlations}} \tabularnewline	
 	War injuries & War captivity & Displacement\tabularnewline
	1.000  & -- & -- & \multicolumn{1}{l}{War injuries}  \tabularnewline
	-0.107 & 1.000  &  -- & \multicolumn{1}{l}{War captivity} \tabularnewline
	0.041 & 0.014 & 1.000 & \multicolumn{1}{l}{Displacement}  \tabularnewline
	\bottomrule
	\end{tabular}
	\end{footnotesize}
	\begin{tablenotes}[flushleft] \item \footnotesize{\emph{Notes}: Panel (a) reports the frequency of each possible combination of treatments in our sample. Panel (b) shows the correlation between treatments. The statistics refer to men born in 1919-21.}
	\end{tablenotes}
	\end{threeparttable}
\end{center}
\end{table}
 
One concern is that soldiers can be subject to multiple war-related shocks. If the different shocks were correlated, we would risk capturing the consequences of multiple war experiences, rather than one specific shock. Appendix Table \ref{tab:corrshocks} shows that this is not a concern in our setting. Panel (a) reports the frequency of each possible combination of treatments, confirming that many individuals are hit by multiple shocks. For example, many displaced soldiers also experienced war captivity or war injuries. However, as shown in panel (b), displacement is not correlated with either imprisonment or wartime injuries. Imprisonment and wartime injuries show a slight negative correlation, presumably because wartime injuries reduced subsequent combat time and thus the risk of imprisonment. However, the correlation is small ($\rho=-0.11$), and Table \ref{tab:cs_robustness191921} shows that our estimates change little when adding the other respective shocks as controls (i.e., when considering all shocks jointly).  
 
Still, one might wonder if the different shocks interact in some way, and how this would affect the interpretation of our estimates. For example, war injuries might have a different effect on displaced than non-displaced soldiers. To understand the implications of such interaction effects for the interpretation of our estimates, imagine there are two treatments, war injuries ($inj=1$) and displacement ($dis=1$). The focus on just two treatments simplifies the exposition. The probability limit of the slope coefficient in an OLS regression of an outcome $y$ on an indicator for war injuries can then be expressed as
 \begin{align} 
 \beta_{\text {OLS }}  = & E[y \mid inj=1]-E[y \mid inj=0] \\ 
  = & E[E[y \mid inj=1, dis]]-E[E[y \mid inj=0, dis] \\ 
  = & \quad \operatorname{Pr}(dis=0 \mid inj=1) E[y \mid inj=1, dis=0] \notag  \\ 
     & +\operatorname{Pr}(dis=1 \mid inj=1) E[y \mid inj=1, dis=1] \notag \\ 
     & -\operatorname{Pr}(dis=0 \mid inj=0) E[y \mid inj=0, dis=0] \notag \\ 
     & -\operatorname{Pr}(dis=1 \mid inj=0) E[y \mid inj=0, dis=1]  \\ 
  = & \quad \operatorname{Pr}(dis=0)(E[y \mid inj=1, dis=0]-E[y \mid inj=0, dis=0]) \notag  \\ 
     & +\operatorname{Pr}(dis=1)(E[y \mid inj=1, dis=1]-E[y \mid inj=0, dis=1])  \label{eq_weights}
 \end{align}
where in the second step we used the law of iterated expectations, and the last step follows because the treatments are uncorrelated, so that $Pr(dis=d | inj=0)=Pr(dis=d | inj=1)=Pr(dis=d)$ for $d={0,1}$. Equation (\ref{eq_weights}) clarifies that our approach captures the weighted average of the treatment effect of war injuries on soldiers who were not displaced ($E[y | inj=1,dis=0]-E[y | inj=0,dis=0]$) and those who were displaced ($E[y | inj=1,dis=1]-E[y | inj=0,dis=1]$). The weights correspond to the respective population shares of each group. If war injuries have different effects on displaced and non-displaced soldiers (\textit{heterogeneous treatment effects}), we thus identify their effect on the ``average person'' in our sample. Alternatively, if the treatment effect of war injuries does not vary between displaced and non-displaced soldiers (\textit{homogenous treatment effects}), then the fact that some individuals are affected by multiple treatments would be irrelevant for the interpretation of our estimates.
 
One might want to go a step further, and separately identify heterogeneous treatment effects, such as the effect of war injuries on individuals who were not displaced ($E[y \mid inj=1, dis=0]-E[y \mid inj=0, dis=0]$), and the effect on those who were displaced ($E[y \mid inj=1, dis=1]-E[y \mid inj=0, dis=1]$). While understanding the heterogeneity underlying our average treatment effects would be interesting, the sample size for some of these comparisons would be very small; for example, there are only 33 soldiers in our sample who experienced a war injury but were neither displaced nor in war captivity. For our baseline analysis, we therefore prefer to exploit the full variation of war exposures in the sample, to estimate an appropriately (population-share) weighted average effect of each war shock. 

\begin{figure}[!tb]
	\caption{Life-cycle effects of war experiences, heterogeneity of treatment effects}\label{fig:lfeffects_het}
	\centering
	\begin{threeparttable}
		\begin{center}
			\textit{\begin{center}
					War injury\\ \smallskip
			\end{center}}
			\subfloat[Employment]{\includegraphics[clip, trim=0.3cm 0.7cm 0.3cm 0.7cm, width=0.48\textwidth]{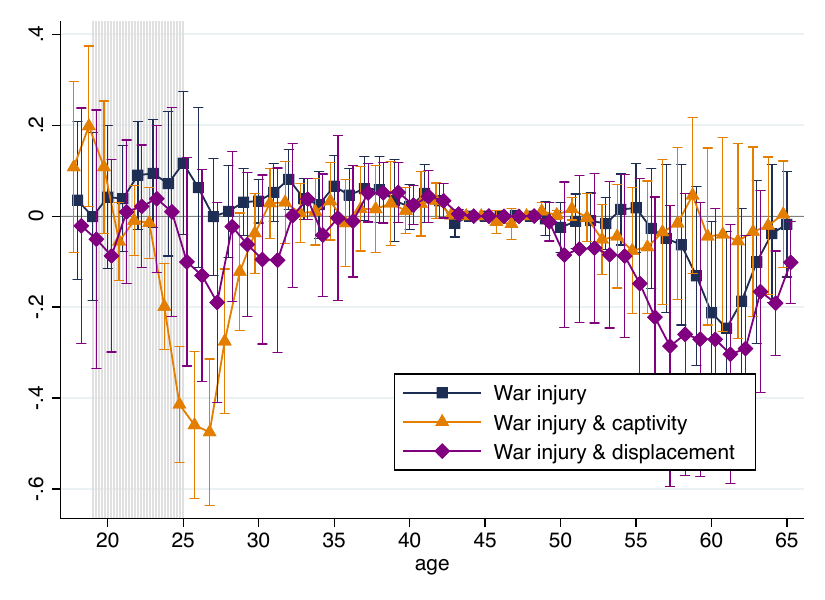}}
			\hskip 7pt
			\subfloat[Occupational Prestige]{\includegraphics[clip, trim=0.3cm 0.7cm 0.3cm 0.7cm, width=0.48\textwidth]{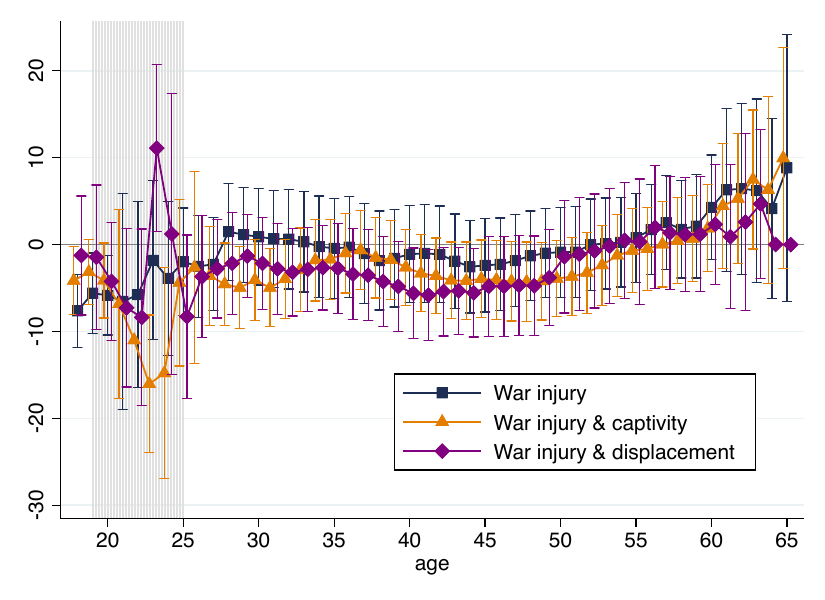}} \\
			\textit{\begin{center}
					War captivity\\ \smallskip
			\end{center}}
			\subfloat[Employment]{\includegraphics[clip, trim=0.3cm 0.7cm 0.3cm 0.7cm, width=0.48\textwidth]{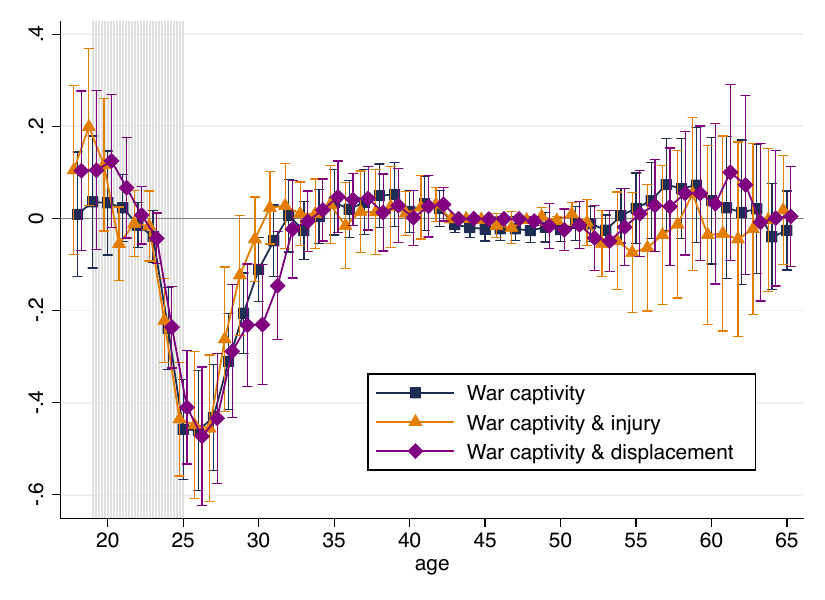}}
			\hskip 7pt
			\subfloat[Occupational prestige]{\includegraphics[clip, trim=0.3cm 0.7cm 0.3cm 0.7cm, width=0.48\textwidth]{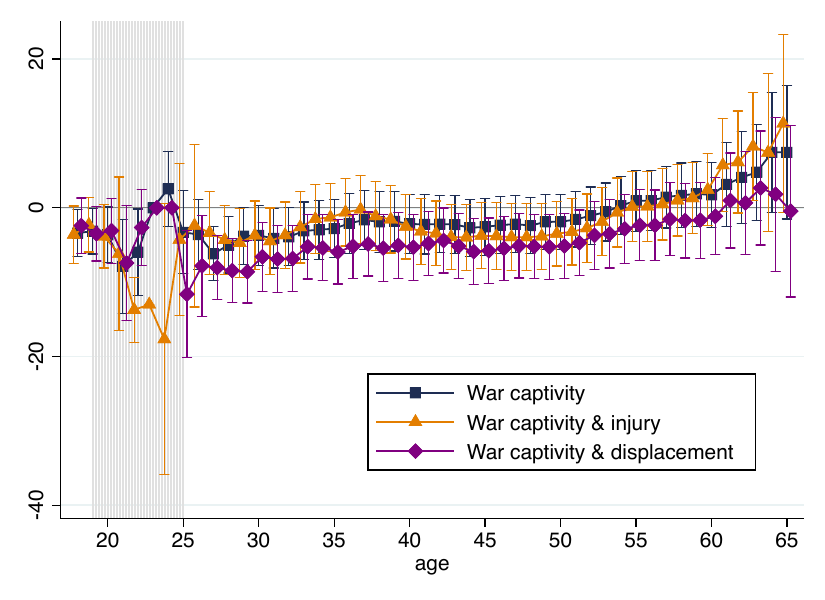}}\\
						\textit{\begin{center}
					Displacement\\ \smallskip
			\end{center}}
			\subfloat[Employment]{\includegraphics[clip, trim=0.3cm 0.7cm 0.3cm 0.7cm, width=0.48\textwidth]{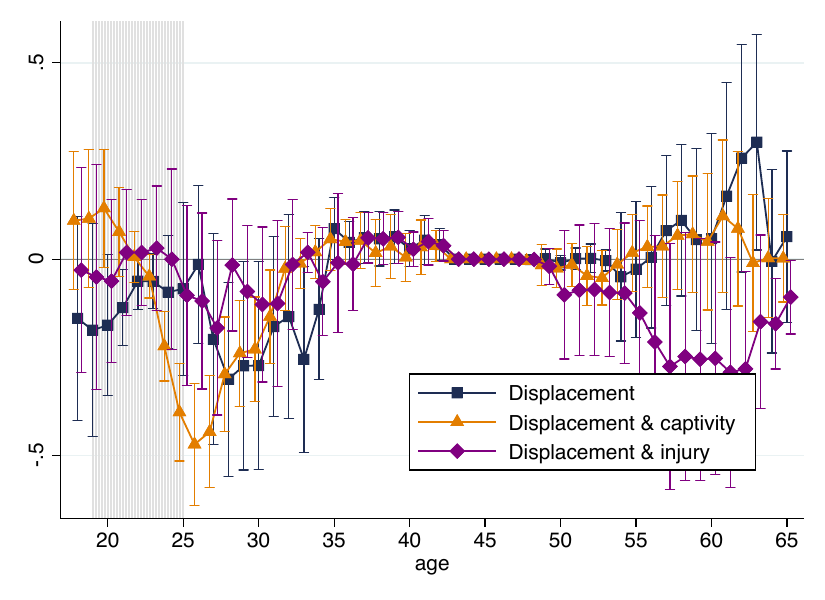}}
			\hskip 7pt
			\subfloat[Occupational prestige]{\includegraphics[clip, trim=0.3cm 0.7cm 0.3cm 0.7cm, width=0.48\textwidth]{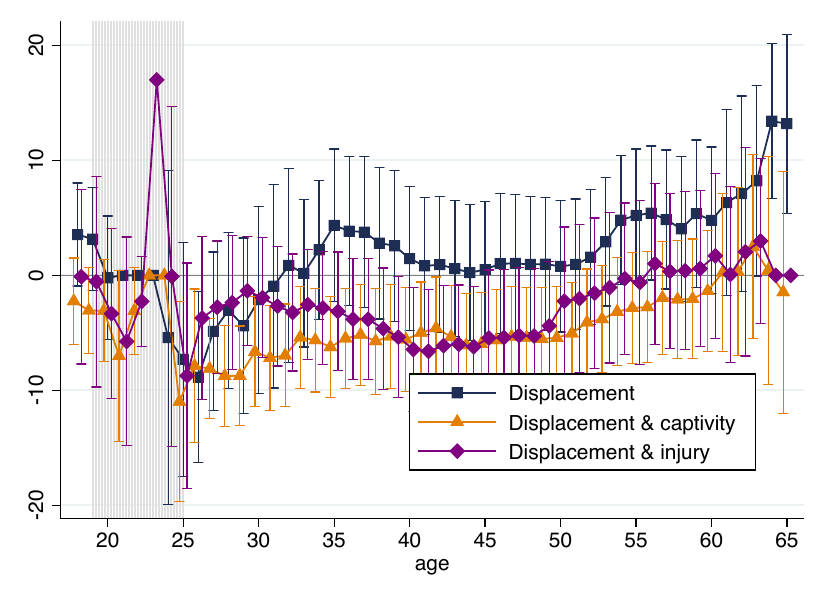}}
		\end{center}
		\begin{tablenotes}[flushleft]
			\item \footnotesize{\emph{Notes}: Heterogeneous effects of war-related injuries, war captivity, and displacement on employment (left panels) and occupational prestige (conditional on employment, right panels) over the life cycle. Estimates are from a pooled OLS regression, interacting indicators for the treatment group of interest and birth year with a full set of age indicators. The sample consists of males born 1919-21, and the control group was not exposed to any war-related shock. The vertical bands indicate the 95\% confidence interval of each estimate. The shaded area indicates the duration of WWII.}
		\end{tablenotes}
	\end{threeparttable}
\end{figure}

Still, to show that our main findings hold also in pairwise comparisons, we revisit our evidence on the employment effects over the lifecycle (Figure \ref{fig:lfeffects_191921}) to study combinations of treatment effects. The estimates are shown in Figure \ref{fig:lfeffects_het}. For each figure, we consider four different groups, a control group that was not exposed to any of our three war-related shock, and three treatment groups that were exposed to different combinations of treatments. For example, in panels (a) and (b) we estimate separate effects for those who experienced a war injury but no other war-related shocks, those who experienced injury and captivity, and those who experienced injury and displacement. 

These subgroup-specific estimates are substantially more noisy, but otherwise confirm the patterns from our baseline analysis as shown in Figure \ref{fig:lfeffects_191921}. For example, soldiers experiencing a war injury have similarly high employment rates as the control group over most of their life, but are more likely to leave employment at older ages. We see this decline of old-age employment both for injured veterans who were displaced (purple line), and those who were not (black line), but not for those injured veterans who also experienced war captivity -- consistent with our findings that war captivity tends to increase employment at older age. More generally, the estimated coefficient for double treatments as shown in Figure \ref{fig:lfeffects_191921} tend to be similar to the combined effect of each as reported in our main analysis, although some are imprecisely estimated.

\section{Additional Figures and Tables}
\subsection{Figures}
\begin{figure}[h]
	\caption{German territorial losses in World War I and II}
	\label{fig:TerritorialLosses}
	\begin{threeparttable}
		\begin{center}
			\includegraphics[width=0.75\textwidth, angle=360]{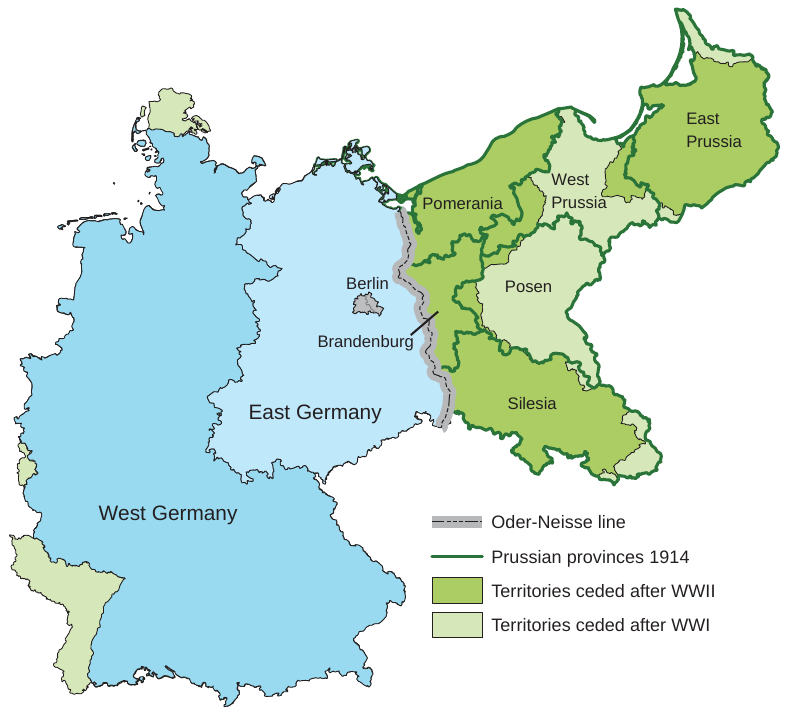}
		\end{center}
		\vspace{0.5cm}
		\begin{tablenotes}[flushleft]
			\item \footnotesize{\emph{Base maps}: \cite{MPI2011}.}
		\end{tablenotes}
	\end{threeparttable}
\end{figure}

\begin{figure}[ht!]
	\caption{The effect of war injury or captivity on health over the life cycle, cohort 1919-21}
	\label{fig:injury_illness}
	\centering
	\begin{threeparttable}
		\begin{center}
			\subfloat[War injury]{\includegraphics[width=0.48\textwidth]{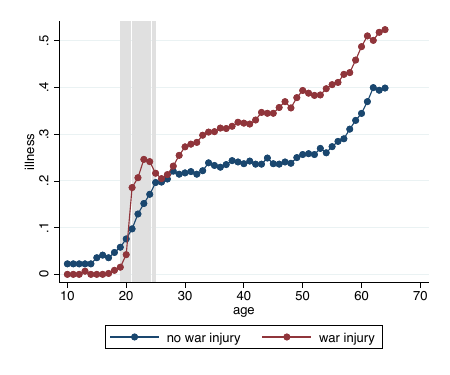}}
			\hfill
			\subfloat[War captivity]{\includegraphics[width=0.48\textwidth]{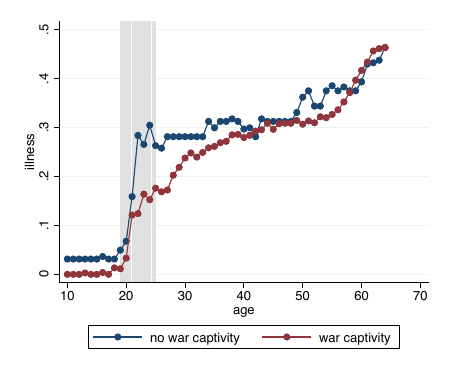}}
		\end{center}
		\vspace*{-1mm}				
		\begin{tablenotes}[flushleft]
			\item \footnotesize{\emph{Notes}: The figure plots the share of respondents reporting ill health over the life cycle, comparing those who sustained a war-related injury (left panel) or those who experienced war captivity (right panel) with those who did not. The sample consists of males born 1919-21 who served in the war. The shaded area indicates the duration of WWII.}
		\end{tablenotes}
	\end{threeparttable}
\end{figure}

\medskip 

\begin{figure}[ht!]
	\caption{The effect of displacement on education over the life cycle, cohort 1929-31}
	\label{fig:displ_education}
	\centering
	\begin{threeparttable}
		\begin{center}
			\subfloat[Males]{\includegraphics[width=0.48\textwidth]{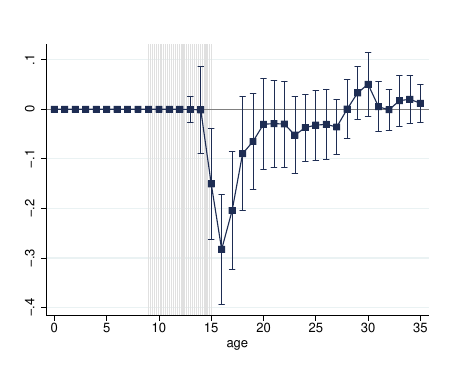}}
			\hfill
			\subfloat[Females]{\includegraphics[width=0.48\textwidth]{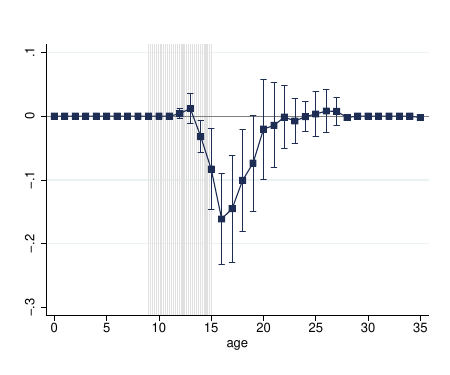}}
		\end{center}
		\vspace*{-1mm}		
		\begin{tablenotes}[flushleft]
			\item \footnotesize{\emph{Notes}: The graph depicts estimated differences in educational participation between displaced and non-displaced individuals over the life cycle, drawing on GHS data. Estimates come from conditional OLS regressions, controlling for the father's and mother's years of schooling and number of siblings. Point estimates are marked by a dot. The vertical bands indicate the 95\% confidence interval of each estimate. The shaded area indicates the duration of WWII.}
		\end{tablenotes}
	\end{threeparttable}
\end{figure}

\clearpage

\subsection{Tables}

\begin{table}[!ht]
	\centering
	\begin{center}
		\begin{threeparttable}
			\caption{Exogeneity of war service and war shocks (bivariate)} \centering
			\label{tab:exoshocksbi}
			\begin{footnotesize}	
			\begin{tabular}{lcccc}
			\toprule 
			 & War service & War injury & War captivity & Displaced \\
			 & (0/1) & (0/1) & (0/1) & (0/1) \\
			 & men & men & men & men \& \\
			 & only  & only &  only & women \\
			 & (1) &(2) &(3)&(4)\tabularnewline
			\midrule
			\multicolumn{4}{l}{\textbf{Pre-war characteristics}} \tabularnewline
	        		Father's years of schooling & -0.002 & -0.013 & 0.007 & -0.003 \\
       			 & (0.006) & (0.012) & (0.011) & (0.006) \\
       			Mother's years of schooling & -0.012 & -0.025 & 0.016 & 0.003 \\
        			& (0.011) & (0.024) & (0.022) & (0.013) \\
        			Father's occupational score & -0.001 & 0.000 & -0.002 & 0.000 \\
        			& (0.001) & (0.002) & (0.002) & (0.001) \\
        			Birth year  & 0.014 & -0.014 & 0.009 & 0.008 \\
        			& (0.012) & (0.026) & (0.023) & (0.015) \\
        			\# siblings & -0.005 & 0.006 & 0.005 & 0.006 \\
        			& (0.004) & (0.009) & (0.008) & (0.005) \\
        			Years of schooling & 0.001 & -0.024* & 0.003 & 0.016* \\
        			& (0.006) & (0.013) & (0.012) & (0.008) \\
        			Poor health before age 18 & -0.273*** & -0.050 & -0.083 & -0.096** \\
        			& (0.036)	& (0.091) & (0.082) & (0.042) \\
        			Female & & & & 0.003 \\ 
        			& & & & (0.024) \\				
			\bottomrule
			\end{tabular}
			\end{footnotesize}
			\begin{tablenotes}[flushleft] \item \footnotesize{\emph{Notes}: The table reports coefficient estimates from bivariate regressions of the indicated war-related shock on each pre-war individual and parental characteristics for birth cohorts 1919-21. Estimates for war injuries and captivity are conditional on war service. Standard errors in parentheses. ***, **, and * denote statistical significance at the 1\%, 5\%, and 10\% level, respectively.}
			\end{tablenotes}
		\end{threeparttable}
	\end{center}
\end{table}

\begin{table}[t!]
	\centering
	\begin{threeparttable}
		\caption{The effect of war captivity duration and location} \centering
		\label{tab:captivity_continuous}
		\begin{footnotesize}
			\begin{tabular}{lccccccc}
				\toprule
				&Educational& \multicolumn{2}{c}{Years in employment} & Occup.  &  \multicolumn{3}{c}{Old age income from}  \\ \cline{3-4} \cline{6-8}
				&attainment&  \hspace{0.4cm}age\hspace{0.4cm}  & \hspace{0.3cm}age\hspace{0.3cm} & prestige  & work  & war victim & nonpension\\
				&(years)& 20-55 & 56-65 & (maximum) & pension  & pension & sources \\
				& (1)	& (2)	& (3)  & (4) & (5) & (6) & (7)  \\
				\midrule
				\multicolumn{4}{l}{\textbf{(a) War captivity (in years)}}  & & & & \\
			 	Raw	&      -0.135** &      -0.792***&      -0.060   &      -0.595** &     -35.16   &     -20.89***&     -15.38   \\
           			 	&     (0.053)   &     (0.078)   &     (0.068)   &     (0.278)   &    (31.44)   &     (6.34)   &    (30.24)   \\
				Baseline&      -0.082*  &      -0.815***&      -0.072   &      -0.400   &     -21.60   &     -20.09***&      -1.32   \\
            				&     (0.049)   &     (0.079)   &     (0.071)   &     (0.284)   &    (33.93)   &     (6.64)   &    (33.09)   \\
				Extended&      -0.057   &      -0.779***&      -0.043   &      -0.336   &     -14.66   &     -21.42***&      11.87   \\
            				&     (0.042)   &     (0.080)   &     (0.072)   &     (0.266)   &    (36.92)   &     (7.13)   &    (35.74)   \\
				\multicolumn{6}{l}{\textbf{(b) War captivity (in years), interacted with location of captivity}}  & & \\
				War captivity &      -0.164*  &      -0.433***&       0.093   &      -0.184  & - & - & -  \\
            			\, (in years) &     (0.089)   &     (0.104)   &     (0.156)   &     (0.656)   \\
				East/USSR x &       0.131   &      -0.714***&      -0.344** &      -0.099  & - & - & - \\
            			\, War captivity  &     (0.104)   &     (0.235)   &     (0.167)   &     (0.783)   \\	
				\bottomrule
			\end{tabular}%
		\end{footnotesize}
		\begin{tablenotes}[flushleft]
			\item \footnotesize{\emph{Notes}: Estimates of the effect of war captivity duration on various outcome variables. Each estimate is from a separate regression. The sample consists of males born 1919-21. In panel (a), the ``raw'' specification controls only for birth year indicators. The ``baseline'' specification additionally controls for years of schooling of father and mother, number of siblings, and time of entry into the war. The ``extended'' specification additionally controls for own years of secondary schooling, an indicator for poor health before age 18, and all other war shocks. Panel (b) corresponds to the ``baseline'' specification but interacts war captivity with an indicator for being interned in Eastern Europe (incl. USSR), which is observed only in the first part of the GHS-2 (N=153). Pension outcomes in columns (5) to (7) are observed only in the second part of GHS-2 conducted (see Footnote \ref{fn:lvs}). Robust standard errors in parentheses, ***, **, and * denote statistical significance at the 1\%, 5\%, and 10\% level.}
		\end{tablenotes}
	\end{threeparttable}
\end{table}

\begin{table}[t!]
	\centering
	\begin{threeparttable}
		\caption{The effect of displacement by prewar SES} \centering
		\label{tab:displacement_prewarSES}
		\begin{footnotesize}
			\begin{tabular}{lccccccc}
				\toprule
				&Educational& \multicolumn{2}{c}{Years in employment} & Occup.  &  \multicolumn{3}{c}{Old age income from}  \\ \cline{3-4} \cline{6-8}
				&attainment&  \hspace{0.4cm}age\hspace{0.4cm}  & \hspace{0.3cm}age\hspace{0.3cm} & prestige  & work  & war victim & nonpension\\
				&(years)& 20-55 & 56-65 & (maximum) & pension  & pension & sources \\
				& (1)	& (2)	& (3)  & (4) & (5) & (6) & (7)  \\
				\midrule
				\multicolumn{3}{l}{\textbf{Displacement (0/1)}}  & & & & \\
				Displacement  &      -0.229   &      -0.595   &       0.175   &      -2.556** &     -78.039   &      18.749   &    -330.914***\\
           			 &     (0.248)   &     (0.362)   &     (0.309)   &     (1.031)   &   (147.677)   &    (40.637)   &    (95.582)   \\
 				Displ. x Father's &      -0.403   &       0.218   &       0.275*  &      -0.674   &      20.863   &     -10.125   &    -147.669** \\
            			\, schooling &     (0.256)   &     (0.162)   &     (0.142)   &     (0.857)   &   (109.514)   &    (13.103)   &    (62.433)   \\
				Observations          &     427   &     427   &     427   &     427   &     254   &     254   &     269   \\
				\bottomrule
			\end{tabular}%
		\end{footnotesize}
		\begin{tablenotes}[flushleft]
			\item \footnotesize{\emph{Notes}: Estimates of the effect of displacement on various outcome variables. Each estimate is from a separate regression. The sample consists of males born 1919-21. ***, **, and * denote statistical significance at the 1\%, 5\%, and 10\% level, respectively.}
		\end{tablenotes}
	\end{threeparttable}
\end{table}

\begin{table}[t!]
   \centering
	\begin{threeparttable}
		\caption{Life-cycle effects of war-related shocks, men born 1919-21} \centering
		\label{tab:lifecycle}
	\begin{footnotesize}
		\begin{tabular}{lccccc}
			\toprule
			& \multicolumn{5}{c}{Age brackets} \\
			\cmidrule{2-6}
			&18--25&26--30&31--40&41--55&56--65 \\
			& (1)		& (2)		& (3) 		& (4) 		& (5) 	\\
			\midrule
			\multicolumn{3}{l}{\textbf{(a) War injury (0/1)}}  & & \\
			In education&--0.017&--0.023&--0.008&-&- \tabularnewline
			&(0.011)&(0.017)&(0.012)&& \tabularnewline
			Employed&0.028**&0.022&0.004&--0.015&--0.085*** \tabularnewline
			&(0.014)&(0.030)&(0.011)&(0.011)&(0.029) \tabularnewline
			Occupational Score&--1.760**&--1.629*&--0.263&--0.337&0.770 \tabularnewline
			&(0.890)&(0.975)&(0.964)&(1.067)&(1.209) \tabularnewline
			Married&0.006&0.088**&--0.022&--0.004&0.008 \tabularnewline
			&(0.016)&(0.042)&(0.024)&(0.010)&(0.018) \tabularnewline
			Children&0.007&0.106&--0.004&--0.135&--0.135 \tabularnewline
			&(0.022)&(0.073)&(0.101)&(0.131)&(0.137) \tabularnewline
			\midrule
			\multicolumn{3}{l}{\textbf{(b) War captivity ($>$ 6 months)}}  & & \\
                        	In education&--0.033**&--0.000&0.023**&-&- \tabularnewline
                        	&(0.015)&(0.018)&(0.011)&& \tabularnewline
                        	Employed&--0.082***&--0.282***&--0.003&--0.008&0.043 \tabularnewline
                        	&(0.018)&(0.029)&(0.015)&(0.010)&(0.033) \tabularnewline
                        	Occupational Score&--2.344**&--4.280***&--2.458**&--2.366*&--1.658 \tabularnewline
                        	&(0.955)&(1.251)&(1.237)&(1.276)&(1.539) \tabularnewline
                        	Married&--0.021&--0.232***&--0.013&0.015&0.015 \tabularnewline
                        	&(0.020)&(0.047)&(0.030)&(0.015)&(0.024) \tabularnewline
                        	Children&--0.024&--0.203**&--0.110&0.112&0.133 \tabularnewline
                        	&(0.034)&(0.085)&(0.115)&(0.148)&(0.155) \tabularnewline
			\midrule
			\multicolumn{3}{l}{\textbf{(c) Displacement (0/1)}}  & & \\
                        	In education&--0.006&0.007&0.010&-&- \tabularnewline
                        	&(0.013)&(0.020)&(0.013)&& \tabularnewline
                        	Employed&0.007&--0.080**&--0.012&--0.008&0.011 \tabularnewline
                        	&(0.014)&(0.036)&(0.011)&(0.014)&(0.031) \tabularnewline
                        	Occupational Score&--1.228&--4.801***&--3.081***&--3.497***&--3.315*** \tabularnewline
                        	&(0.996)&(0.996)&(1.025)&(1.051)&(1.237) \tabularnewline
                        	Married&--0.001&--0.076&--0.032&--0.015&--0.017 \tabularnewline
                        	&(0.019)&(0.048)&(0.031)&(0.014)&(0.024) \tabularnewline
                        	Children&0.022&0.006&--0.064&--0.218&--0.212 \tabularnewline
                        	&(0.032)&(0.084)&(0.119)&(0.152)&(0.161) \tabularnewline
			\bottomrule 
		\end{tabular}%
	\end{footnotesize}
		\begin{tablenotes}[flushleft]
			\item \footnotesize{\emph{Notes}: Estimates of the effect of war-related shocks on various outcomes (shown on the left) at different ages (shown in the table header). The sample consists of males born 1919-21. All regressions control for birth year (indicators), years of schooling of father and mother, number of siblings and time of entry into the war (all interacted with age). The regressions for occupational prestige are estimated conditional on being employed. Robust standard errors in parentheses, ***, **, and * denote statistical significance at the 1\%, 5\%, and 10\% level.}
		\end{tablenotes}
	\end{threeparttable}
\end{table}

\begin{table}[!t]
	\centering
	\begin{threeparttable}
		\caption{The effect of displacement on education, by sex, cohort, and data source} \centering
		\label{tab:displacement_educ_LVS}
		\begin{footnotesize}
			\begin{tabular}{lccccccc}
				\toprule
				& \multicolumn{3}{c}{Males} & & \multicolumn{3}{c}{Females} \\
				\cmidrule{2-4}  	\cmidrule{6-8} 
				&1919-21 & 1929-31 & 1939-41 & &1919-21 & 1929-31 & 1939-41 \\
				& (1)	 & (2)	   & (3) & & (4)	 & (5)	   & (6)	 \\
				\midrule
				\multicolumn{8}{l}{\textbf{(a) GHS}}\\
				Displacement (0/1) 	& -0.150	& -0.652***  & 0.074 & 	& 0.360* & -0.825*** & 0.207   \\
				& (0.233)  	& (0.274) 	 & (0.343) &	& (0.202) & (0.242) 	&(0.403)   \\
				& [10.934] & [10.693] &[10.782] &	& [9.732] 	& [9.447]	&[9.988] \\
				Observations		& 427 & 303 & 310 &			& 605 & 303 & 298 \\
				\midrule
				\multicolumn{8}{l}{\textbf{(b) 1970 census}}\\ 
				Displacement (0/1) 	& -0.091***	  	& -0.626*** 	& 0.217*** & 	& 0.137***	  		& -0.369*** & 0.207***   \\
				& (0.025)  		& (0.023) 	&(0.021) && (0.018) & (0.017) 	&(0.019)   \\
				& [11.390] 		& [11.431]	&[11.860] && [9.452] 		& [9.501]	&[10.366] \\
				Observations		& 82,980 & 104,279 & 138,095 && 119,309 & 103,202 & 133,548 \\
				\bottomrule	
			\end{tabular}
		\end{footnotesize}
		\begin{tablenotes}[flushleft]
			\item \footnotesize{\emph{Notes}: The table shows, by sex and cohort, estimates of the effect of displacement on years of education, drawing on data from the GHS (panel (a)) and 1970 census (panel (b)). Years of education include time spent in vocational training and at university. Estimates come from conditional OLS regressions. All regressions control for birth year (indicators). The GHS regressions additionally control for years of schooling of father and mother, number of siblings and (for males) time of entry into the war. Robust standard errors in parentheses, unconditional means for the non-displaced control group in square brackets. ***, **, and * denote statistical significance at the 1\%, 5\%, and 10\% level.}
		\end{tablenotes}
	\end{threeparttable}
\end{table}

\clearpage
\makeatletter
\setlength{\@fptop}{0pt plus 1fil}
\setlength{\@fpbot}{0pt plus 1fil}
\makeatother

\section{Theoretical Predictions}\label{sec:theory} 

How do war-related shocks affect an individuals' education and labor market outcomes over the life course, and can standard theory capture those effects? In this section we derive theoretical predictions from a standard life-cycle model of human capital and retirement decisions.

\subsection{A Ben-Porath model with endogenous retirement}

Summarizing a version of the Ben-Porath model with endogenous retirement decisions \citep{Hazan2009}, assume that an individual's lifetime utility $V$ equals
\begin{equation}
	V = \int_{0}^{T}e^{-\rho t}u(c(t))dt-\int_{0}^{R}e^{-\rho t}f(t)dt,
\end{equation}
where $c(t)$ is consumption at age $t$, $f(t)$ is the disutility of work (assumed to satisfy $f^{\prime}(t)>0$ and $f(T)=\infty$), $\rho$ is the subjective discount rate, $R$ is the retirement age, and $T$ is the length of the individual's lifetime.

Human capital $h(s)$ and therefore the wage $w$ depend on the individual's choice of the length of schooling prior to entering the labor market $s$ and ``learning speed'' $\theta(s)$, such that $w=h(s)=e^{\theta(s)}$. The sole costs of schooling is foregone earnings, so the budget constraint 
\begin{equation}
	\int_{s}^{R}e^{-rt}e^{\theta(s)}dt=\int_{0}^{T}e^{-rt}c(t)dt
\end{equation}
equates consumption over the lifetime (between 0 and $T$) with earnings over the working life (between $s$ and $R$), where $r$ is the interest
rate. Following \cite{Hazan2009}, we assume $r=\rho$, implying that consumption is constant over the life cycle, 
\begin{equation}
	\ensuremath{c(s,R)}=\frac{e^{\theta(s)}\left(e^{-rs}-e^{-rR}\right)}{1-e^{-rT}}.\label{eq:opt_consumption}
\end{equation}
Solving the Lagrangian associated with maximizing lifetime utility $V$ 
leads to the two equilibrium conditions equating the marginal costs of schooling with its marginal benefits,
\begin{equation}
	\ensuremath{\frac{1}{\theta^{\prime}(s)}=\frac{1-e^{-r(R-s)}}{r}}\label{eq:EQ1}
\end{equation}
and the disutility of work at age $R$ with the marginal utility of working (in terms of consumption)\begin{equation}\ensuremath{f(R)=u^{\prime}(c(s,R))e^{\theta(s)}}.\label{eq:EQ2}
\end{equation}
Figure \ref{fig:theory}a provides a numerical example, assuming $u(\cdot)=log(\cdot)$, $f(R)=1/(1-R)$, $T=r=1$ and $\theta(s)=s/0.3$. The (thin) blue line corresponds to the indifference curve associated with the optimal schooling condition in equation (\ref{eq:EQ1}) while the (thick) orange line corresponds to the optimal retirement decision represented by equation (\ref{eq:EQ2}). The optimal schooling and retirement age are determined by the intersection of these two curves (point A). 

\begin{figure}[!tbp]
	\caption{Theoretical predictions}\label{fig:theory}
	\centering
	\begin{threeparttable}
		\begin{center}
			\subfloat[Baseline]{\includegraphics[width=0.45\textwidth]{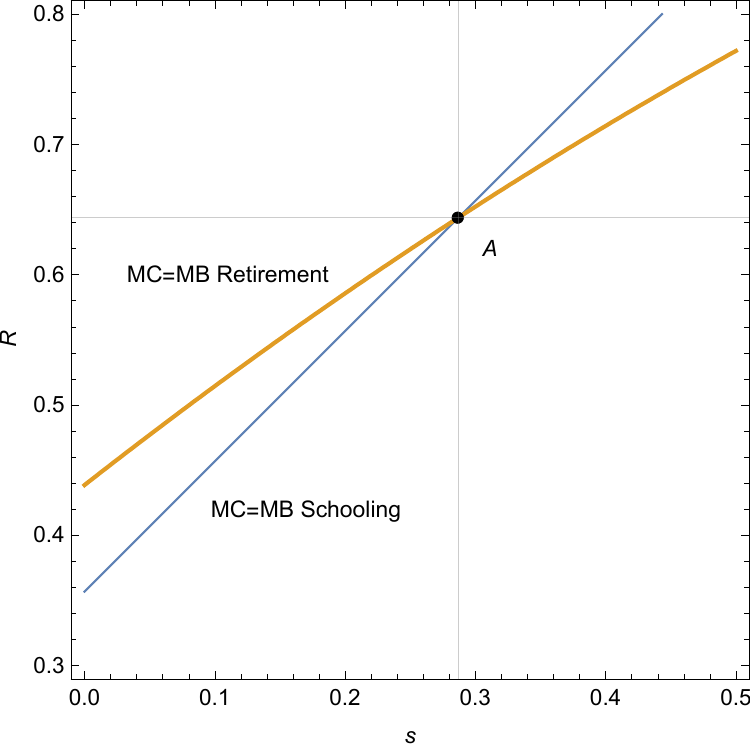}}\hfill
			\subfloat[War-related injury]{\includegraphics[width=0.45\textwidth]{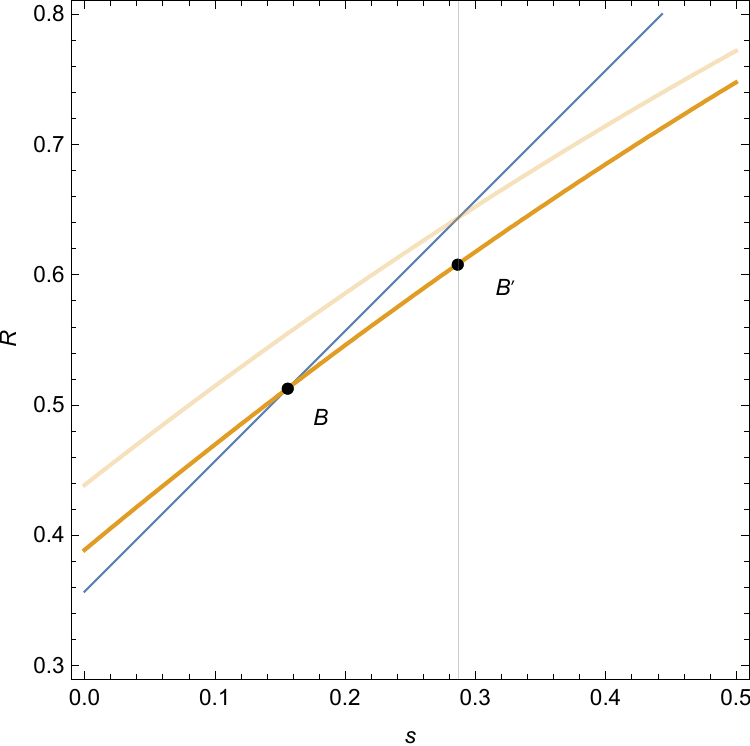}}\hfill \\
			\subfloat[War captivity]{\includegraphics[width=0.45\textwidth]{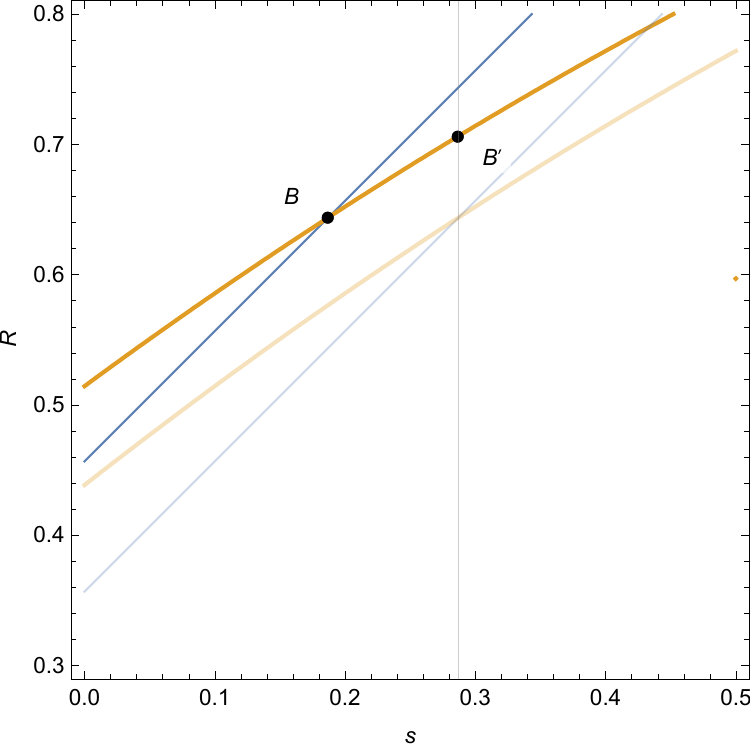}}\hfill
			\subfloat[Displacement]{\includegraphics[width=0.45\textwidth]{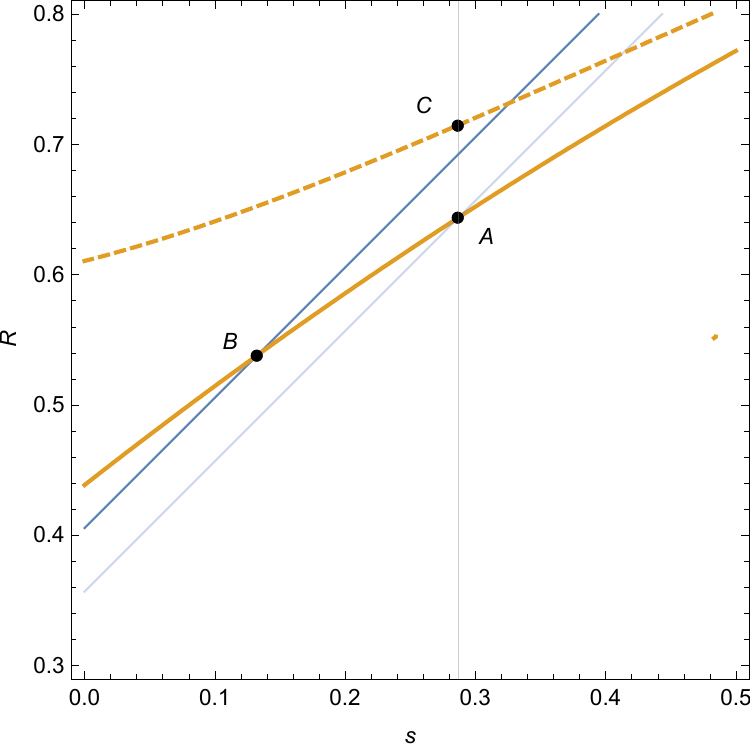}}\hfill						
		\end{center}
		\begin{tablenotes}[flushleft]
			\item \footnotesize{\emph{Notes}: Numerical illustrations of a Ben-Porath model with retirement decision \citep{Hazan2009} for retirement age $R$ and length of schooling $s$. Figure (a) is our baseline calibration with $u(\cdot)=log(\cdot)$, $f(R)=1/(1-R)$, $T=r=1$ and $\theta(s)=s/0.3$. Figure (b) corresponds to a war-related injury with an increase in disutility of work such that $f_{injury}(R)=1.2 f(R)$. Figure (c) corresponds to war captivity with time $x=0.1$ spent in captivity. Figure (d) corresponds to displacement with a reduction in the wage rate to $\theta_{d}(s)=0.9\theta(s)$ (solid blue and orange lines) or a reduction in wealth such that $c_{d}(s,R)=c(s,R)-1/3$ (dashed orange line).}
		\end{tablenotes}
	\end{threeparttable}
\end{figure}

Using this model, we next derive the implications of different types of war-related shocks--war injuries, captivity, and displacement--for the choice of schooling $s$ and retirement age $R$.

\subsection{War injuries}\label{sec:theory_injuries} 
Among men born 1919-21, nearly one third suffered injuries such as bullet and shrapnel wounds, frostbite or amputations (see Table \ref{tab:prevalenceshocks}). 
What are the likely implications? Interpreted through the lens of the model, war injuries increase the disutility of work $f(R)$. Moreover, their effect on schooling $s$ will be non-positive, as explained below. The equilibrium condition (\ref{eq:EQ2}) determining the retirement decision then implies that the marginal utility of consumption $u^{\prime}$ must increase, corresponding to a reduction in consumption (and income), and therefore a reduction in the retirement age $R$.\footnote{War injuries might also reduce the length of the individual's lifetime $T$, reducing the retirement age $R$ further. The reason is that according to condition (\ref{eq:opt_consumption}), a decrease in $T$ increases the consumption level (for a given $R$ and $s$). Consequently,  the marginal utility of consumption decreases, and so does retirement age $R$ (according to the condition (\ref{eq:EQ2})). However, this mechanism is less relevant in our context, as our empirical analysis conditions on survival until the statutory retirement age.}

But, war injuries may not only affect the disutility of work, but also generate different types of income effects. First, the war-injured were eligible to a war pension (see Section \ref{subsec:pensions}), corresponding to an \emph{increase} in income. Second, war-related injuries may reduce productivity, at least in some jobs or for some individuals, corresponding to a decrease in the wage $w=e^{\theta(s)}$ and a \emph{decline} in income and pensions. The overall effect of war injuries on income is therefore ambiguous. We show in Section \ref{sec:results1} that in our setting, the net effect on (labor + war) pensions is negligible. The main channel via which war injuries affect (life-cycle) income is therefore the retirement decision.  

The effect of war injuries on schooling $s$ is non-positive. First, note that most of those born in 1919-21 entered the military around age 20, \emph{after} leaving school. Moreover, military service shortened the remaining lifespan available for work, thereby lowering the incentives for war returnees to invest into education (a positive relation between the length of the economic lifespan and educational investments is a standard implication of the Ben-Porath model; see \citealt{BenPorath1967}). Indeed, fewer than 10\% of returnees entered an apprenticeship after the war.
\footnote{While the war may have increased skill returns overall, such general equilibrium effect would affect both the treated (the war-injured) and control group.} A further shortening of the active working life due to early retirement decreases these incentives further, implying that the effect of war injuries on educational investments are negative.\footnote{As a possible exception, educational investments might allow the war-injured to access white-collar jobs, in which war injuries might be less detrimental to productivity than in blue collar jobs. However, we do not find such occupational reallocation in our setting.} 

Figure \ref{fig:theory}b provides a numerical example. The increase in the disutility of work corresponds to a downward shift of the indifference curve associated with condition (\ref{eq:EQ2}). If the war-injured could freely optimize (ex-ante optimization) they would reduce both retirement entry $R$ and their schooling $s$ (point B). However, most have completed their schooling investments before enlistment to the military (vertical line). As they cannot reduce those educational investments ex-post, their incentives to reduce the retirement age are mitigated (point B'). Standard theory therefore predicts that war injuries decrease the retirement age and reduce educational investments in the right tail of the distribution (i.e., among those who had not yet completed their investments before enlistment).  

\subsection{War captivity}\label{sec:theory_captivity} 
More than three quarters of men born 1919-21 were in captivity, often for years (see Table \ref{tab:prevalenceshocks}). This captivity disincentives educational investments. While some war returnees entered apprenticeships or studied at a university, those spending time in captivity returned later and would have made such investments later. But a key implication of the Ben-Porath and similar models is that educational investments are less profitable at later ages, when the remaining productive work span is shorter. Formally, the optimal educational investment of war prisoners is determined by
\begin{equation}
	\ensuremath{\frac{1}{\theta^{\prime}(s)}=\frac{1-e^{-r(R-x-s)}}{r}}
\end{equation}
where $x$ is the time spent in captivity. An increase in $x$ decreases the right hand of this equation, so for the condition to hold we require a reduction of schooling $s$ or an increase in the retirement age $R$ (or both). 

Individuals choose their optimal retirement age according to condition (\ref{eq:EQ2}). Plausibly, the disutility of work on the left-hand side is not much affected by war captivity. But for a given retirement age $R$ and schooling $s$, the right side of (\ref{eq:EQ2}) increases because life-cycle income--and therefore consumption according to equation (\ref{eq:opt_consumption})--declines due to war captivity.\footnote{As an individual's work-span is shorter than his lifespan, years spent in war captivity will decrease life-cycle income by a greater proportion than the period over which consumption needs to be financed. The effect on pensions will be more modest, as the pension system compensated for gaps in the employment biography due to war captivity (see Section \ref{subsec:pensions}).} Specifically, the optimal consumption is now given by
\begin{equation}
	\ensuremath{c(s,R)}=\frac{e^{\theta(s)}\left(e^{-r(s+x)}-e^{-rR}\right)}{1-e^{-rT}}.
\end{equation} 
Therefore, consumption decreases and the marginal utility of consumption increases in $x$, ceteris paribus. To satisfy condition (\ref{eq:EQ2}) we therefore need that the retirement age $R$ increases and/or that $s$ declines in response to time spent in captivity $x$.\footnote{While an increase in retirement age increases both sides of equation (\ref{eq:EQ2}), it will ultimately increase the left side more (as $f(T)=\infty$).} The reduction in lifetime income associated with captivity therefore raises incentives to work; this is akin to income effects from shifts in non-labor income or wages in standard models of labor supply. 

Figure \ref{fig:theory}c provides a numerical example. Both the indifference curves associated with condition (\ref{eq:EQ1}) and condition (\ref{eq:EQ2}) shift upward, reflecting a decrease in the marginal benefits of schooling for a given level of $R$ and an increase in the marginal utility of working for a given level of $s$. If individuals could freely optimize they would reduce schooling but do not change their retirement age much (point B). 
However, many individuals will have already completed their schooling investments before enlistment (vertical line). With education above its ex-post optimum, individuals have an incentive to retire later (point B'). Standard theory therefore predicts that war captivity increases the retirement age but reduces educational investments and, therefore, wages.

\subsection{Displacement}\label{sec:theory_displacement} 
More than one fifth of our survey respondents are displaced Germans, mostly from the German Reich's Eastern territories (see Table \ref{tab:prevalenceshocks}). The extent to which displacement affects educational and labor market careers will depend on the timing of the expulsion. As most displacements occurred towards the end of WWII, they will have only limited effects on the educational investments of older cohorts, including the 1919-21 cohort. In contrast, younger cohorts experienced direct interruptions of their educational careers. For example, the 1929-31 cohort were only 14-16 years olds when the war ended in 1945. As we show in Section \ref{sec:section5}, displacement therefore led to a large decline in education among younger cohorts. 

Here we focus instead on the labor market effects of displacement. Motivated by the evidence shown in Section \ref{sec:section5}, we assume that displacement reduces an individual's wage from $\theta(s)$ to $\theta_{d}(s)$, such that $\theta_{d}(s)<\theta(s)$ $\forall s$. The precise reason for this wage decline is not central for our argument, but it might reflect the loss of social networks, specific human capital or ``search capital'' as the displaced could not return to their previous jobs.\footnote{See also \cite{Bauer2013}, who show that in 1971, first-generation displaced men had 5.1\% lower incomes than native men and displaced women 3.8\% lower incomes than native women. Moreover, the displaced were markedly over-represented among blue-collar workers and under-represented among the self-employed.} This wage decline affects the marginal utility of working $\ensuremath{u^{\prime}(c(s,R))e^{\theta(s)}}$ on the right-hand side of equilibrium condition (\ref{eq:EQ2}) via two channels. On the one hand, a reduction in the wage $w=e^{\theta(s)}$ directly reduces the incentives to work (\textit{substitution effect}). On the other hand, a reduction in earnings also reduces consumption $c(s,R)$, thereby increasing the marginal benefits of consumption $u^{\prime}(c(s,R))$ and incentives to work (\textit{income effect}). As these income and substitution effects have opposing signs, the overall effect on the optimal retirement age $R$ is ambiguous and depends on the curvature of the utility function. A pure wealth effect on the other hand would generate an income but no substitution effect, and therefore lead to an unequivocal postponement of retirement entry. 

Figure \ref{fig:theory}d provides a simple illustration for the case of displacement \textit{before} labor market entry (such as for the 1919-21 cohort). A proportional decrease in the wage rate $w=e^{\theta(s)}$ due to expulsion would shift the indifference curve associated with condition (\ref{eq:EQ1}) upwards (thin blue line). The curve associated with condition (\ref{eq:EQ2}) however remains unchanged in this particular example: as utility is log linear in consumption we have $u^{\prime}(c(s,R))=1/c(s,R)$, and income and substitution effects of a reduction in the wage rate in early life  cancel each other out exactly. Ex-ante, expellees would therefore choose lower schooling and an earlier retirement age (point B). However, if educational investments were made before expulsion (as for the 1919-21 cohort) the optimal retirement age remains unchanged (point A). If expulsions affect wealth rather than wages then the curve associated with condition (\ref{eq:EQ2}) would shift, corresponding to a pure income effect (orange dashed line). Assuming schooling is fixed at $s$, individuals would choose a higher retirement age (point C). 

But how would the employment effects of displacement vary with age at displacement $d$? The key insight is that the size of the income effect from a reduction in the wage rate depends on the age at which an individual experiences this change. Individuals who are already close to their expected retirement age experience only a minor income effect, as most of their life-cycle earnings have already been realized. The effect of displacement on the employment of older individuals is therefore dominated by the substitution effect, and hence negative. We can show this explicitly by solving for the consumption profile of displaced individuals. Recall that for given life-cycle earnings, the optimal consumption profile is flat. However, as displacement was unexpected, it shifts an individual's consumption from $c$ to post-displacement consumption $c_d$. Focusing on displacement events that occur before retirement ($d<R$) but after the completion of schooling ($d>s$), the budget constraint of a displaced individual is therefore given by
\begin{equation}
\int_{s}^{d}e^{-rt}e^{\theta(s)}dt+\int_{d}^{R_d}e^{-rt}e^{\theta_{d}(s)}dt=\int_{0}^{d}e^{-rt}cdt+\int_{d}^{T}e^{-rt}c_{d}dt
\end{equation}
where the left-hand side is the sum of present discounted value of earnings at wage $e^{\theta(s)}$ before displacement (between end of schooling $s$ and displacement $d$) and at wage $e^{\theta_{d}(s)}$ after displacement (between $d$ and new retirement age $R_d$), and the right side is the PDV of consumption $c$ in the pre- and consumption $c_d$ in the post-period. Solving the budget constraint for $c_d$ and simplifying, we have 
\begin{align}
c_{d} & =\frac{e^{\theta(s)}\left(e^{-rd}-e^{-rs}\right)+e^{\theta_{d}(s)}\left(e^{-rR_{d}}-e^{-rd}\right)-c\left(e^{-rd}-1\right)}{e^{-rT}-e^{-rd}}\nonumber \\
 & =c+\left(e^{\theta_{d}(s)}-e^{\theta(s)}\right)\frac{e^{-rd}-e^{-rR}}{e^{-rd}-e^{-rT}}+\left(e^{-rR}-e^{-rR_d}\right)\frac{e^{\theta(s)}}{e^{-rd}-e^{-rT}}.
\end{align}
The difference between consumption after displacement $c_d$  and consumption $c$ in the pre-period (which itself is a function of schooling $s$ and the planned retirement age $R$) depends on the difference between the old and new wage rate, weighted by the relative lengths of the post-displacement working ($R-d$) and consumption spells ($T-d$). For example, if displacement occurs at the end of the working life ($d=R$), then the weight equals zero and consumption remains unaffected; the individual experiences only a substitution effect. In contrast, if displacement occurs earlier in the career ($R-d\gg0$) then the weight will be closer to one, and the decline in consumption can be substantial; the individual experiences an income effect. This reduction in consumption maps into a corresponding increase in the marginal utility, counteracting the substitution effect from a decline in wages. Consumption post-displacement also depends on the gap between the initially planned and new post-displacement retirement ages, $R$ and $R_d$; an unplanned early retirement due to displacement might necessitate a sudden and large reduction in consumption. 

\begin{figure}[!tbp]
	\caption{Theoretical predictions: Displacement by age}\label{fig:displacement_by_age}
	\centering
	\begin{threeparttable}
		\begin{center}
		\includegraphics[width=0.55\textwidth]{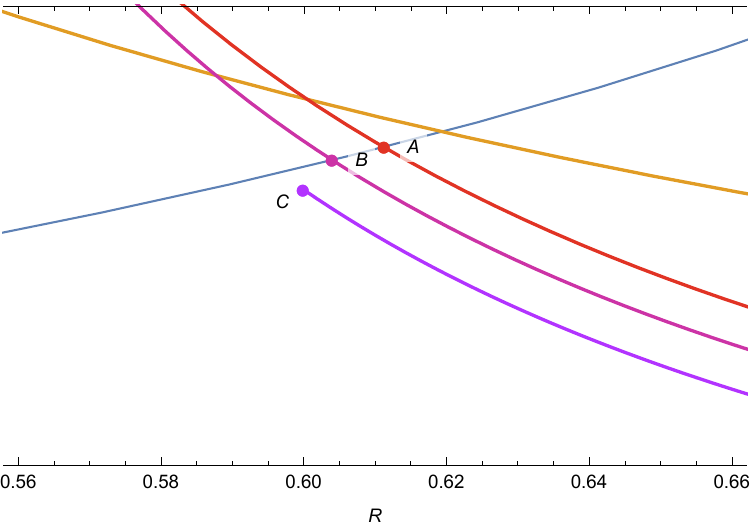}\hfill			
		\end{center}
		\begin{tablenotes}[flushleft]
			\item \footnotesize{\emph{Notes}: Numerical illustrations of a Ben-Porath model with retirement decisions \citep{Hazan2009} with $u(\cdot)=log(\cdot)$, $f(R)=1/(1-R)$, $T=r=1$, $\theta(s)=4s$ and $\theta_{d}(s)=3s$. The thin blue line corresponds to the the disutility of work $f(R)$. The orange line corresponds to the marginal utility of working $\ensuremath{u^{\prime}(c(s,R))e^{\theta(s)}}$.  The red, red-purple and purple lines correspond to the marginal utility of working if an individual is displaced at age $d=0.5, d=0.55$ or $d=0.6$.}
		\end{tablenotes}
	\end{threeparttable}
\end{figure}

Figure \ref{fig:displacement_by_age} illustrates these arguments by plotting the two sides of the equilibrium condition (\ref{eq:EQ2}) for the optimal retirement decision, over retirement age $R$. The marginal disutility of work $f(R)$ increases (thin blue line) while the marginal benefits $\ensuremath{u^{\prime}(c(s,R))e^{\theta(s)}}$ decreases over $R$ (orange line). The optimal retirement age corresponds to the intersection of the two curves. The effects of displacement depend on the age at displacement $d$. We compare displacement around mid-age ($d=0.5$) or closer to retirement age ($d=0.6$). For the mid-age worker, the marginal benefits of working change (red line), but they remain greater than the marginal disutility of work at the time of displacement. This individual would therefore not retire immediately, but retire earlier than originally planned (at point A). Intuitively, mid-aged expellees cannot "afford" to leave the labor force immediately, as their cumulative life-cycle earnings are still low. They retire earlier, but not much earlier than planned. For a worker displaced at older age ($d=0.6$), this income effect is less pronounced, and their response to displacement is instead dominated by the substitution effect. Indeed, in our example the marginal benefit of working (purple line) falls below the marginal disutility of work; the individual would therefore retire immediately after displacement (point C). Point B represents an intermediate case, in which displacement occurs at $d=0.55$. Overall, a simple life-cycle model implies that the immediate effect of displacement on employment exit increases with age-at-displacement, in line with the empirical pattern shown in Figure \ref{fig:displ_empl_exit}c).

A similar mechanism could explain why the displacement effect increases less steeply with age-at-displacement for women than men. At the time of our study, most women were married and rarely the main breadwinner in the household. The marginal benefits of working will therefore tend to be lower but also flatter over retirement age $R$ for women (as consumption is less dependent on own than on the spouse's employment). This has two implications. First, young women may leave the labor force in response to displacement, while this is unlikely to be the case for young men. Second, the employment effects of displacement are reduced for older women, as many will have left the labor force already. 
 
\begin{small} 

\end{small}

\end{document}